\documentclass[amsmath,amssymb,amsbsy,reprint,pra,preprintnumbers,showpacs,superscriptaddress]{revtex4-2}
\usepackage{graphicx,color}
\usepackage{dcolumn}
\usepackage{bm}
\usepackage{braket}
\usepackage{mathtools}
\usepackage{ulem}
\usepackage[breaklinks,colorlinks=true,linkcolor=blue,urlcolor=blue,citecolor=blue]{hyperref}
\usepackage{times}

\begin{document}
\title{Blocking particle dynamics in diamond chain with spatially increasing flux}
\author{Tomonari Mizoguchi}
\affiliation{
Department of Physics,
University of Tsukuba, Tsukuba, Ibaraki 305-8571, Japan
}
\email{mizoguchi@rhodia.ph.tsukuba.ac.jp}
\author{Yoshihito Kuno}
\affiliation{
Graduate School of Engineering Science, Akita University, Akita 010-8502, Japan
}
\author{Yasuhiro Hatsugai}
\affiliation{
Department of Physics,
University of Tsukuba, Tsukuba, Ibaraki 305-8571, Japan
}

\date{\today}

\begin{abstract}
Spatial non-uniformity in tight-binding models
serves as a source of rich phenomena. 
In this paper, we study a diamond-chain tight-binding model with a spatially-modulated magnetic flux at each plaquette.
In the numerical studies with various combinations of the minimum and maximum flux values, 
we find the characteristic dynamics of a particle, namely, 
a particle slows down when approaching the plaquette 
with $\pi$-flux.
This originates from 
the fact that the sharply localized eigenstates exist 
around the $\pi$-flux plaquette.
These localized modes can be understood from a squared model of the original one.
This characteristic blocked dynamics will be observed
in photonic waveguides or cold atoms. 
\end{abstract}

\maketitle

\section{Introduction}
Dynamics of particles in tight-binding models 
has attracted considerable interest.
Particle dynamics contains various useful information about the properties of the systems,
such as the localized nature~\cite{Hatsugai2001} and nontrivial topology~\cite{Mazza2015,Meier2016,Wang2017,Cardano2017,Gong2018,Zhang2019,Maffei2018,Haller2020,Kuno2020,Mizoguchi2021}.
It also provides a novel notion based on the dynamical properties~\cite{Heyl2013,Dora2015,Okugawa2021}. 
Moreover, such dynamical properties have become experimentally accessible.
The tight-binding-type models have originally been 
introduced to describe the electronic structures in solid.
Recently, it has been recognized that tight-binding 
models describe various systems having discrete translational symmetry,
such as ultracold atoms in an optical lattice~\cite{Jaksch1998,Jaksch2003,Maciej2007,Bloch2008}, 
light in photonic waveguides~\cite{Joannopoulos1997,Ozawa2019}, 
and wave motions in mechanical systems~\cite{Kane2014,Ma2019}.

The roles of spatial modulations to Hamiltonians in wave functions and dynamics have also attracted considerable interest.
One of the most well-known phenomena induced by the spatial modulation is Anderson localization~\cite{Anderson1958,Abrahams1979,Evers2008},
where disorders turn extended wave functions into exponentially localized ones.
The drastic change of the wave functions 
is also caused by disorder-free modulations.
For instance, a uniform electric field that causes a linear potential
induces the localization of the wave functions, 
which is called the Wannier-Stark localization~\cite{Wannier1960}.
Then the resulting dynamics becomes 
oscillatory rather than accelerated.
This oscillation of the particle 
dynamics is called Bloch oscillation, 
and is experimentally realized in various artificial setups \cite{Atala2013,Kohlert2023}. 
Recently, the roles of 
the characteristic band structures and 
Bloch wave functions, such as Dirac fermions and flat bands, 
in the aforementioned electric-field-induced phenomena
have been investigated extensively~\cite{Lim2012,Khomeriki2016,DiLiberto2020,Kitamura2020}. 

\begin{figure}[b]
\begin{center}
\includegraphics[clip,width = 0.95\linewidth]{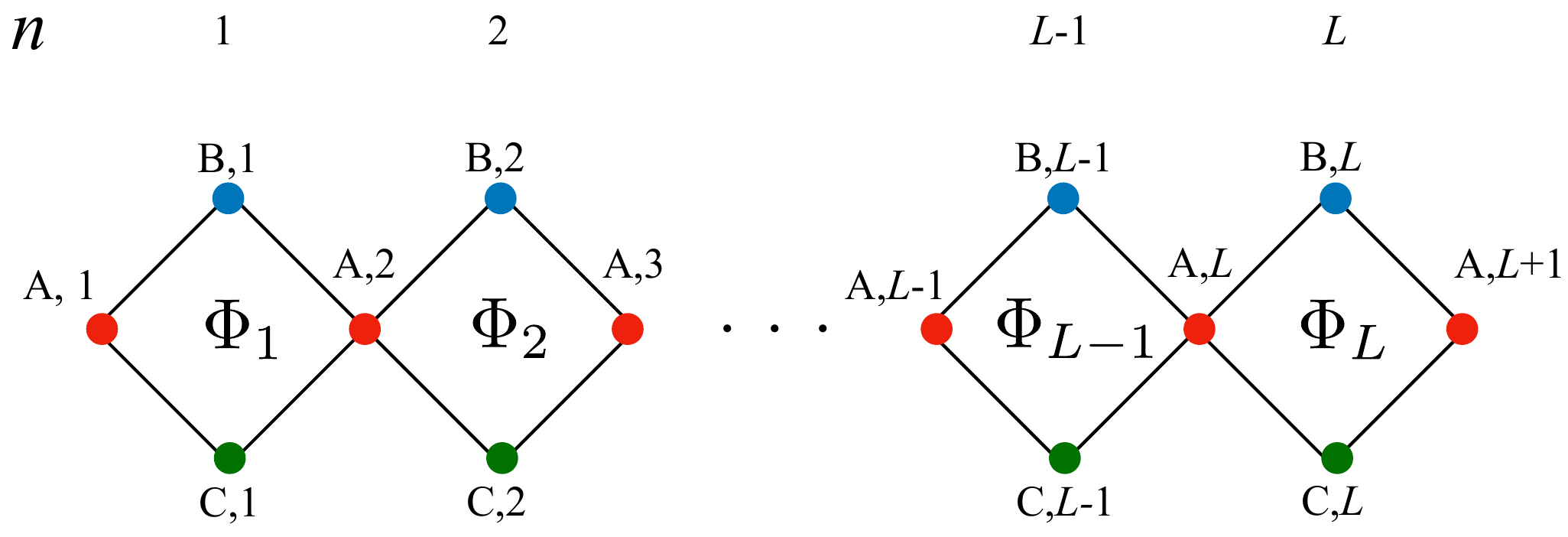}
\vspace{-10pt}
\caption{Schematic figure of the diamond chain model with spatially increasing flux.}
  \label{fig:DC_model}
 \end{center}
 \vspace{-10pt}
\end{figure}

In this paper, we seek another disorder-free modulation of Hamiltonians 
that cause characteristic dynamics of tight-binding models.
Specifically, we introduce the diamond chain model 
with a spatially-increasing flux.
The diamond chain is a one-dimensional, 
corner-sharing network of square plaquettes (Fig.~\ref{fig:DC_model}).
In the tight-binding models on this lattice, 
we can introduce the flux at each plaquette 
as a Pierls phase. 
In fact, the effects of the uniform flux 
in the diamond chain have been studied intensively.
When the flux is equal to $\pi$ (per flux quantum),
all the bands become completely dispersionless, 
resulting in the complete confinement of the particle motion.
Such a flux-induced localization is called an Aharonov-Bohm cage~\cite{Vidal1998,Vidal2000,Vidal2001,Dousot2002,Mosseri2022,Ahmed2022,Kolovsky2023,Marques2023}, 
and is experimentally realized in various setups such as 
photonic crystals~\cite{Mukherjee2018,Kremer2020} and electric circuits~\cite{Zhang2023}.
(We summarize the characteristic band structures for the uniform flux case in Appendix~\ref{app:bulkband}.)
 
In the present work, 
we consider the case where the flux penetrating 
the leftmost (rightmost) plaquette is 
$\Phi_{\rm min}$  $(\Phi_{\rm max})$
and that between them is increased linearly.
We investigate the characteristic 
localization and dynamics of this model.
We first investigate the case of 
$(\Phi_{\rm min}, \Phi_{\rm max}) = (0,\pi)$.
We reveal that the eigenstates can be categorized into several types.
Among them, we find that sharply localized eigenstates near the $\pi$-flux plaquette appear, 
whose energy is close to the finite-energy flat band in the uniform $\pi$-flux case.
We also investigate the single-particle dynamics 
where the particle is initially localized at one or a few sites.
We find a characteristic feature of the wavefront, 
namely, the particle slows down as it approaches the $\pi$-flux plaquette. 
Remarkably, this behavior resembles neither the ballistic motion of the uniform system
nor the Bloch oscillation in the linear potential. 
On the basis of these results, we further study the case of various choices of $\Phi_{\rm min}$ and $\Phi_{\rm max}$.
We find that the localized states around $\pi$-flux plaquette appear ubiquitously,
and such states serve as a blockade of the particle dynamics. 

The rest of this paper is structured as follows.
In Sec.~\ref{sec:model}, we introduce our model and its basics such as symmetries. 
Then, our main results of this work are presented in Sec.~\ref{sec:results}. 
We first study in detail the cases where the flux is increased from 0 to $\pi$ 
and 0 to 2$\pi$.
Based on these results, we discuss particle dynamics of various combinations of $\Phi_{\rm min}$ and $\Phi_{\rm max}$.
Section~\ref{sec:comaprison} is 
devoted to the comparison between the diamond chain model and 
the other models with increasing magnetic flux to 
elucidate the uniqueness and ubiquity of the diamond chain.
We also address another aspect of the 
characteristic dynamics, namely, 
the early time dynamics.  
Finally, we present the summary of this paper 
and several future perspectives in Sec.~\ref{sec:summary}.

\section{Model \label{sec:model}}
We study the tight-binding Hamiltonian:
\begin{align}
H  =&  \sum_{n= 1}^{L}
c^\dagger_{\mathrm{A},n} \left(c_{\mathrm{B},n} +c_{\mathrm{C},n} \right)+ c^\dagger_{\mathrm{A},n+1}
\left(e^{-i\Phi_n}c_{\mathrm{B},n} +c_{\mathrm{C},n} \right)\notag \\
+& (\mathrm{H.c.}),
\end{align}
where $c_{\rm {A/B/C}, n}$ is the annihilation operator
and $\Phi_n$ is a spatial-dependent flux.
Note that we focus on open boundary case and the total number of sites is $N_{\rm site} = 3L+1$.
For convenience, we introduce the matrix representation of this Hamiltonian:
\begin{align}
    H = \hat{\bm{c}}^\dagger \mathcal{H}  \hat{\bm{c}},
\end{align}
where $\hat{\bm{c}}$ is the column vector 
of the annihilation operator and 
$ \mathcal{H} $ is the Hamiltonian matrix. 
$\mathcal{H}$ preserves the chiral symmetry, 
namely, $\mathcal{H}$ satisfies
$ g \mathcal{H} g = -\mathcal{H}$,
with 
\begin{align}
[g]_{ij} = g_{i}\delta_{i,j},
\end{align}
$g_{i} = +1$ ($-1$) for $i \in \mathrm{A}$ ($i \in \mathrm{B,C}$). 

By diagonalizing the Hamiltonian, we have
\begin{align}
H = \sum_{\nu} \varepsilon_\nu  \alpha^\dagger_{\nu} \alpha_{\nu},
\end{align}
where 
\begin{align}
    \alpha_\nu^\dagger = \sum_{i} \psi^{\nu}(i) c^\dagger_i,
\end{align}
is the creation operator of the $\nu$-th eigenstate, 
$\alpha_\nu$ is its Hermitian conjugate, and
$\psi^{\nu}(i)$ is the wave function at the site $i$.

In the following, we study the case of spatially increasing flux. 
Specifically, we set $\Phi_n = \Phi_{\rm min} + \Delta \Phi (n-1)$, $\Delta \Phi := \frac{\Phi_{\rm max}-\Phi_{\rm min}}{L-1}$. This situation induces different strength of flux for each plaquette in the system as shown in Fig.~\ref{fig:DC_model}.

\begin{figure}[t]
\begin{center}
\includegraphics[clip,width = 0.99\linewidth]{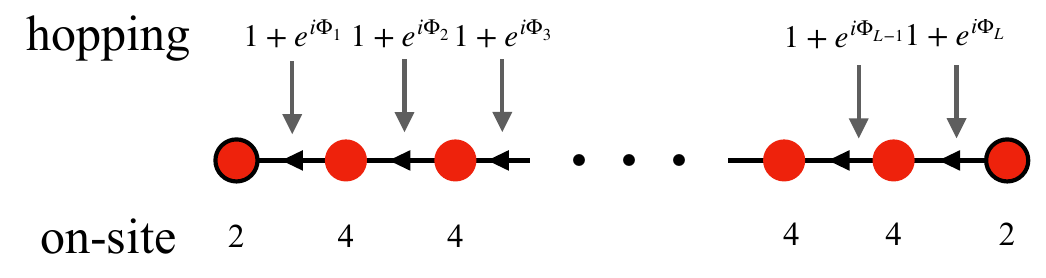}
\vspace{-10pt}
\caption{Schematic figure of the Hamiltonian $h^{\rm A}$.}
  \label{fig:ha}
 \end{center}
 \vspace{-10pt}
\end{figure}
Before proceeding to the numerical results of various $(\Phi_{\rm min},\Phi_{\rm max})$, 
we address the generic properties of the eigenvalues and eigenstates.
Firstly, the chiral symmetry of $\mathcal{H}$ indicates that the positive and negative energy modes appear
in a pairwise manner, and that there exist degenerate zero-energy modes, whose number is equal to $|\mathrm{Tr}(g)| = (L-1)$~\cite{Sutherland1986,Lieb1989,Brouwer2002,Koshino2014}.
In fact, the degenerate zero-energy modes 
are spanned by the compact localized states (CLS), shown in Appendix~\ref{app:CLS}.

Secondly, the chiral symmetry also indicates that 
taking a square of the Hamiltonian provides a perspective on the finite-energy modes~\cite{McClure1956,Arkinstall2017,Attig2017,Kremer2020,Mizoguchi2020_sq,Mizoguchi2021_SRTSM,Yoshida2021,Navarro2023,Matsumoto2023}.
If we align the basis $\hat{\bm{c}}$
as $\hat{\bm{c}}=
\left(c_{\mathrm{A},1},\cdots, c_{\mathrm{A},L+1},
c_{\mathrm{B},1},\cdots, c_{\mathrm{B},L},
c_{\mathrm{C},1},\cdots, c_{\mathrm{C},L}\right)^{\rm T}$,
we can write the Hamiltonian matrix in the following form,
\begin{align}
\mathcal{H} = 
\begin{pmatrix}
\mathcal{O}_{L+1, L+1} &\Omega^\dagger \\ 
\Omega & \mathcal{O}_{2L, 2L} \\
\end{pmatrix},
\end{align}
where 
$\mathcal{O}_{M_1, M_2}$ stands for 
the $M_1 \times M_2$ zero matrix and 
$\Omega$ is the $2L \times (L+1)$ matrix
that describes the hopping between A sites and B/C sites. 
Taking the square of $\mathcal{H}$,
we have 
\begin{align}
\mathcal{H}^2 = 
\begin{pmatrix}
h^{\rm A} & \mathcal{O}_{L+1, 2L}\\ 
\mathcal{O}_{2L, L+1} & h^{\rm B,C} \\
\end{pmatrix},
\end{align}
where $h^{\rm A} :=  \Omega^\dagger \Omega$ and $h^{\rm B,C} :=  \Omega \Omega^\dagger$.
Let $\bm{u}_{\nu}$ be a normalized eigenvector of $h^{\rm A}$ with an eigenvalue $E_{\nu}$.
Since $h^{\rm A}$ is positive semi-definite, $E_{\nu}\geq 0$ holds. 
In the following, we assume that $E_{\nu}>0$.
Then, we find the following two facts:
(i) The vector $\bm{u}_{\nu}^\prime = \frac{1}{\sqrt{E_{\nu}}} \Omega \bm{u}_\nu$ 
is a normalized eigenvector of $h^{\rm B,C}$,
and (ii) the vector $\bm{\psi}^{\pm}_\nu = \frac{1}{\sqrt{2}} \left(\bm{u}_{\nu},  \pm\bm{u}_{\nu}^\prime \right)^{\rm T}$
is an eigenvector of $\mathcal{H}$ with an eigenvalue $\pm \sqrt{E}$~\cite{Mizoguchi2020_sq,Mizoguchi2023,Matsumoto2023}.
The above facts, in combination with an additional fact that 
the matrix elements of $\Omega$ are restricted to pairs of neighboring sites,
indicate the following: 
If $\bm{u}_\nu$ is a sharply localized wave function, so are $\bm{u}^\prime_\nu$ 
and $\bm{\psi}^{\pm}_\nu$.
Let us focus on $h^{\rm A}$, 
which corresponds to a tight-binding Hamiltonian of the $L+1$-site chain.
In Fig.~\ref{fig:ha}, we show the schematic figure of $h^{\rm A}$.
It contains the on-site potentials and the nearest-neighbor hoppings.
In particular, the hopping parameter between the $n$th site and the $(n+1)$th site
is given by $t_{n,n+1} =t^\ast_{n+1,n}  = 1+e^{i\Phi_n}$.
This indicates the following: 
Suppose that there exists a plaquette whose flux value is close to $\pi$, i.e., $\Phi_n = \pi +\delta \Phi$ 
with $\delta \Phi$ being a small number.
Then, the hopping parameter is approximated as $t_{n,n+1} \sim -i \delta \Phi$.
This means that 
the hoppings for $h^{\rm A}$ near $n$ 
corresponding to $\pi$-flux plaquette are largely suppressed, 
which gives rise to a sharply localized eigenstate, $\bm{u}_\nu$ (but, its state is not compact-support due to the small finite contribution of hopping $-i\delta\Phi$ around $\pi$-flux plaquette).
Turning to the diamond chain,
the resulting $\bm{\psi}^{\pm}_\nu$ around the $\pi$-flux plaquette is sharply localized too,
as we shall see in the next section.

\begin{figure}[tb]
\begin{center}
\includegraphics[clip,width = 0.95\linewidth]{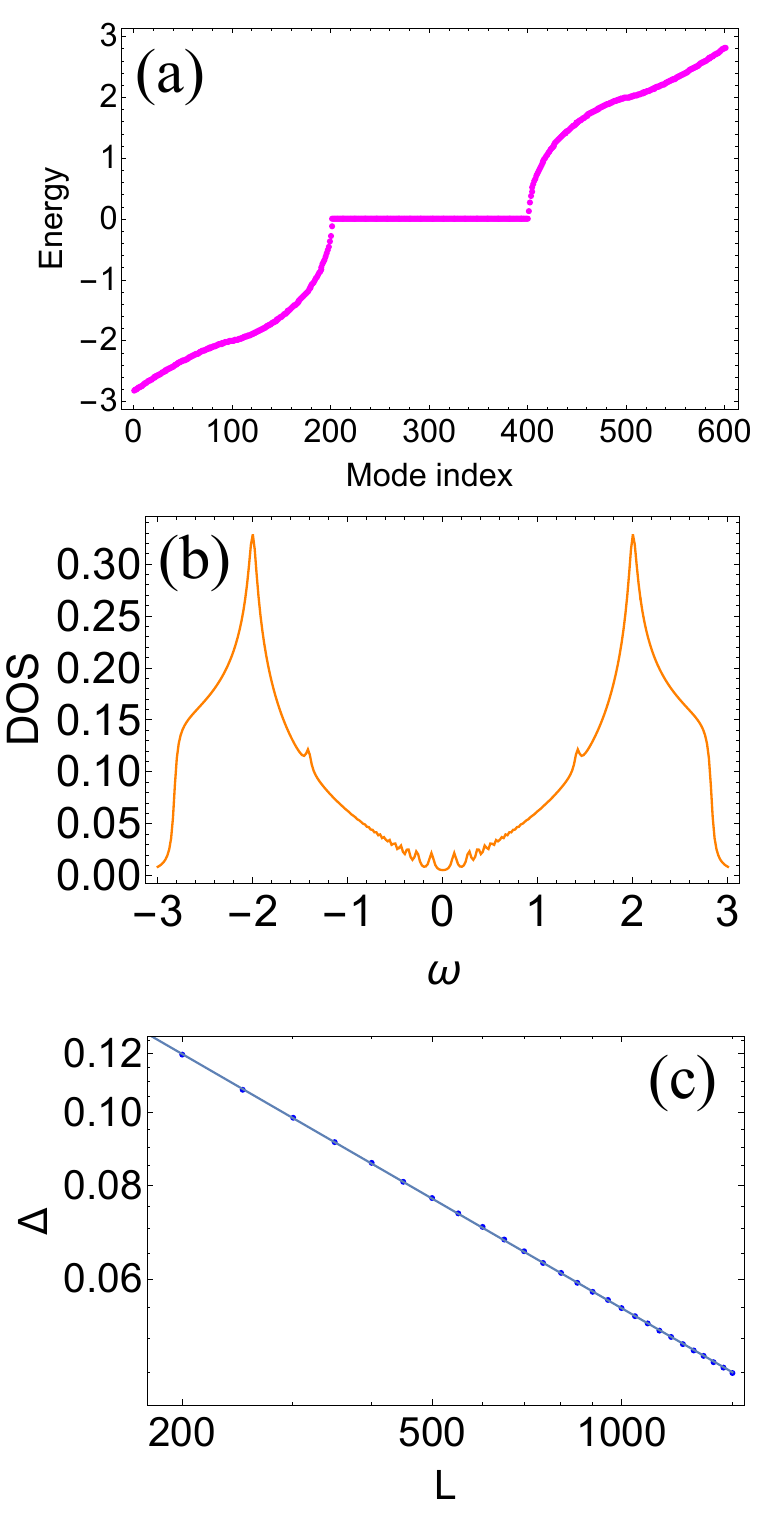}
\vspace{-10pt}
\caption{(a) Energy spectrum and (b) DOS for $(\Phi_{\rm min}, \Phi_{\rm max})=(0,\pi)$, $L=200$.
In (b), the contribution from the degenerate zero-energy modes is excluded.
(c) The gap between the zero-energy mode 
and the lowest positive energy mode, $\Delta$, as a function of $L$.}
  \label{fig:DC_model_eigen}
 \end{center}
 \vspace{-10pt}
\end{figure}

\begin{figure*}[tb]
\begin{center}
\includegraphics[clip,width = 0.95\linewidth]{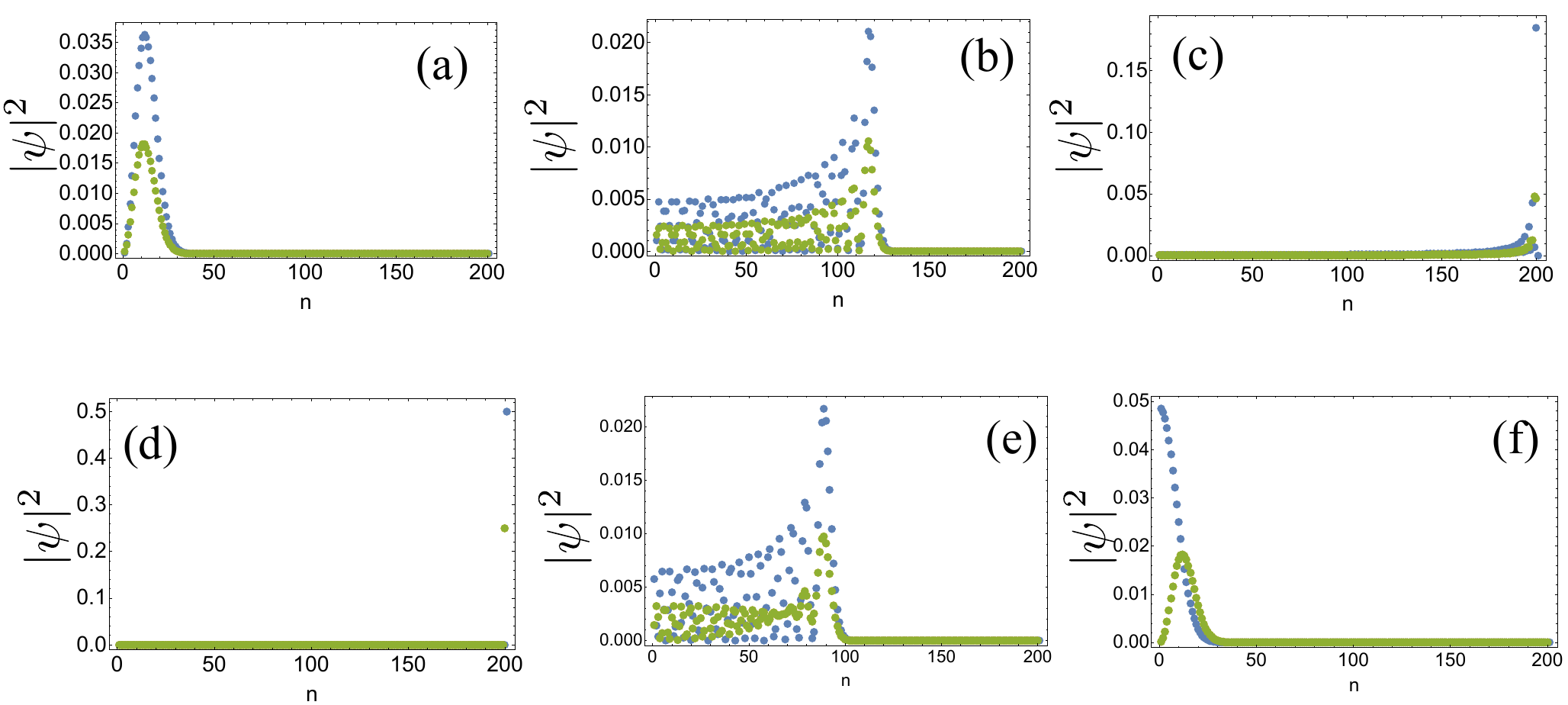}
\vspace{-10pt}
\caption{Probability density distribution, $|\psi^\nu(i)|$, for $(\Phi_{\rm min}, \Phi_{\rm max})=(0,\pi)$, $L=200$ and $\nu=$
(a) 1, (b) 30, (c) 100, (d) 164, (e) 184, and (f) 201.
The blue, orange, and green dots are for the sublattice A, B, and C, respectively (the orange dots overlap the green ones).
The corresponding eigenenergies are (a) $-2.82$,
(b) $-2.52$, (c) $-2.00$,
(d) $-\sqrt{2}$, (e) $-1.00$, and (f) $-0.12$.}
  \label{fig:DC_model_wf}
 \end{center}
 \vspace{-10pt}
\end{figure*}

\section{Results \label{sec:results}}
In this section, we present our numerical results on 
the diamond chain model with various combinations of $(\Phi_{\rm min}, \Phi_{\rm max})$.
\subsection{Case I: $(\Phi_{\rm min}, \Phi_{\rm max})=(0,\pi)$}
We first focus on the case of $(\Phi_{\rm min}, \Phi_{\rm max})=(0,\pi)$.
In fact, the detailed analysis on this case is helpful for understanding the generic cases of $(\Phi_{\rm min}, \Phi_{\rm max})$.
In particular, the role of the $\pi$-flux plaquette is elucidated. 

We first study the energy spectrum 
and energy eigenstates. 
In Fig.~\ref{fig:DC_model_eigen}(a), we plot the energy spectrum.
We see that the zero-energy modes are macroscopically degenerate, as we have mentioned in the previous section. 

In Fig.~\ref{fig:DC_model_eigen}(b), we plot the DOS for the non-zero energy modes, 
defined as
\begin{align}
\mathrm{DOS}(\omega) = - \frac{1}{\pi N_{\rm site}}\sum_{\nu \neq ({\rm zero \hspace{0.5mm} modes}) } \mathrm{Im}\left[\frac{1}{\omega + i\eta -\varepsilon_{\nu}} \right]. \label{eq:DOS}
\end{align}
Here, $\eta$ is a small parameter set as $\eta = 0.03$.
As indicated in Eq.~(\ref{eq:DOS}),
we have excluded the degenerate zero energy modes
that give a divergent contribution to the DOS near $\omega = 0$,
in order to clarify the contribution from non-zero energy modes. 
We see that the DOS drops around $\omega =0$.
We also see the large DOS around $\omega = \pm 2$, which corresponds to the energy of the perfect flat bands for the $\pi$-flux case (see Appendix~\ref{app:bulkband}).

The drop of the DOS around $\omega =0$ raises the question that 
whether the first excited state above
the zero energy modes (i.e., the $2L+1$th mode) has a finite energy gap or not. 
To see this, in Fig.~\ref{fig:DC_model_eigen}(c), we plot the energy gap between the $(2L+1)$th mode and the zero-energy mode, $\Delta$, as a function of $L$. We find that $\Delta$ can be fitted as $\Delta \sim 1.564 \cdot L^{-0.485}$.
Therefore, the spectrum is gapless around the zero-energy.
It is worth noting that, for the uniform flux case, 
the gapless spectrum realizes
only when $\Phi=0$, where the finite energy bands exhibit the Dirac-like linear spectrum (see Appendix~\ref{app:bulkband}).
\begin{figure}[tb]
\begin{center}
\includegraphics[clip,width = 0.95\linewidth]{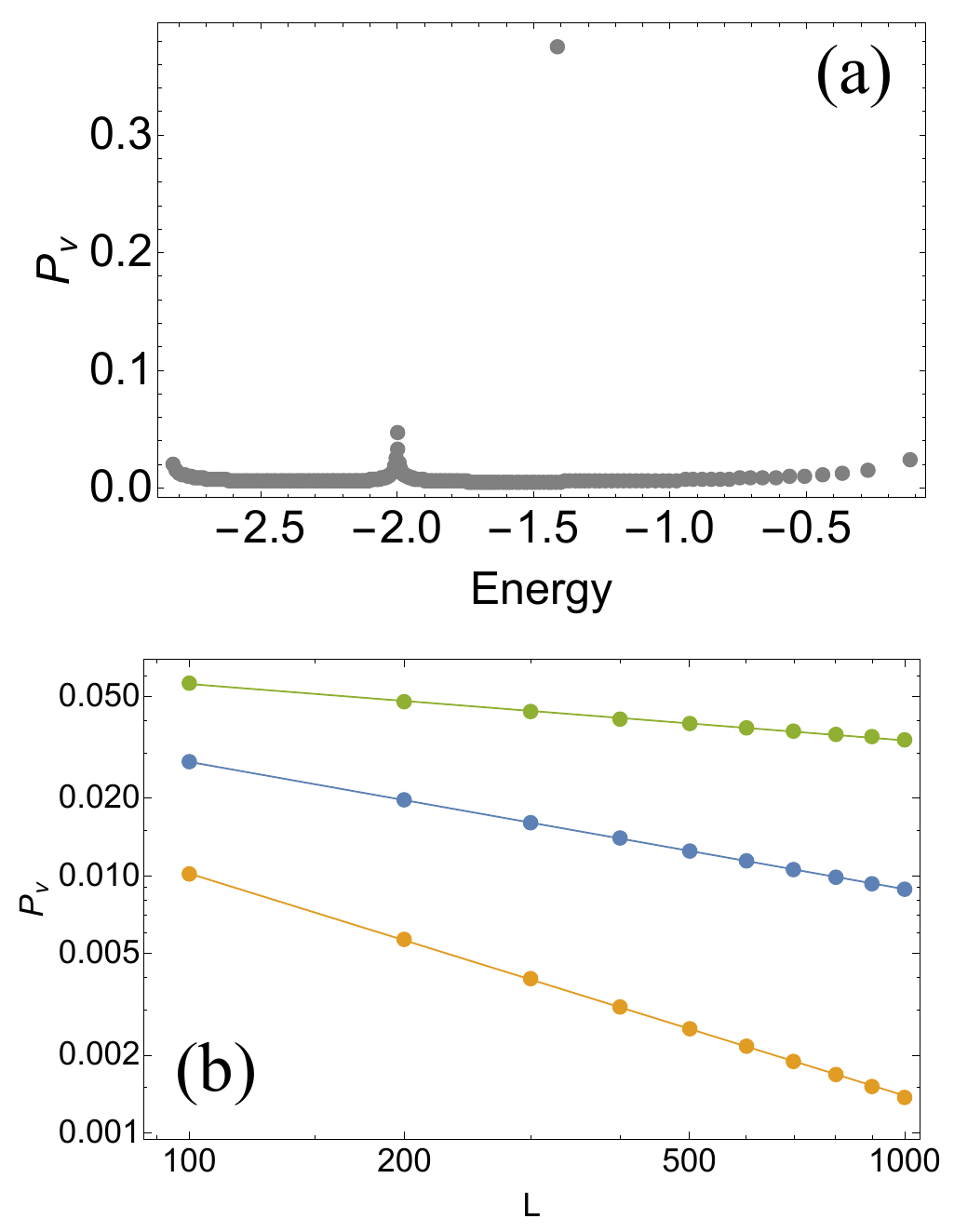}
\vspace{-10pt}
\caption{(a) IPR for the negative energy sector for $L=200$.
(b) $L$ dependence of the IPR for $\nu = 1$ (blue dots), 
$\nu = (3/10) L $ (orange dots) and $\nu =L/2 $ (green dots).
The lines are the fitting curves.}
  \label{fig:ipr_0_pi}
 \end{center}
 \vspace{-10pt}
\end{figure}

We now turn to the features of the wave functions.
In Fig.~\ref{fig:DC_model_wf}, we plot the probability density distribution for some selected values of $\nu$. We focus on $1 \leq \nu \leq L+1$,
i.e., the negative energy sector. 
We find a rich structure of the eigenstates, depending on its eigenenergy.
The most characteristic state is Fig.~\ref{fig:DC_model_wf}(d), where the wave function is compact and localized at the right edge.
In fact, this is the right-edge mode whose eigenenergy is $-\sqrt{2}$ and whose exact wave function is shown in Appendix~\ref{app:right-edge}.
We label the annihilation operator of this edge mode as $R_{-}$;
hereafter, the index $-$ stands for the negative energy sector.
Note that this edge mode is the same as that for the uniform $\pi$-flux case~\cite{Kremer2020}. 
The remaining states are categorized into the following three types:
(i) At the band edge [Figs.~\ref{fig:DC_model_wf}(a) and \ref{fig:DC_model_wf}(f)], 
the wave functions are localized at the left edge. 
We call these modes the band-edge (BE) modes,
and we represent their annihilation operator $X_{\nu,-}$.
(ii) For the state with $E \sim-2$, which corresponds to the finite-energy flat band 
for the uniform $\pi$-flux case (Appendix~\ref{app:finiteCLS}),
the wave functions are sharply localized near the $\pi$-flux plaquette shown in Fig.~\ref{fig:DC_model_wf}(c).
We call these modes the $\pi$-flux-localized (PFL) modes, 
and represent their annihilation operators as $Y_{\nu,-}$.
(iii) The intermediate states have moderate amplitudes on 
the left side of the systems and a vanishing amplitude 
on the right side. The typical probability density distributions of the modes are shown in Figs.~\ref{fig:DC_model_wf}(b) and ~\ref{fig:DC_model_wf}(e). 
We call these modes the intermediate (Int) modes, 
and represent their annihilation operators as $Z_{\nu,-}$.
Note that the origin of the PFL 
modes can be accounted for by the squared Hamiltonian, as we have seen in the previous section.
Also, many of the sharply-localized states 
tend to have a large weight in the right half of the system, since the hopping amplitude for $h^{\rm A}$, $|t_{n,n+1}| = |1+e^{i\Phi_n}|$, 
becomes smaller as $n$ becomes larger.

Summarizing these, we can explicitly write down the structures of the eigenstates as 
\begin{align}
    H =  H_+ + H_-,
\end{align}
where 
\begin{align}
    H_- =& -\sqrt{2} R^{\dagger}_-R_- + \sum_{\nu \in \mathrm{BE}} \varepsilon^{\rm BE}_{\nu} X^\dagger_{\nu,-}X_{\nu,-} \notag \\
    +& \sum_{\nu \in \mathrm{PFL}} \varepsilon^{\rm PFL}_{\nu} Y^\dagger_{\nu,-}Y_{\nu,-}
    + \sum_{\nu \in \mathrm{Int}} \varepsilon^{\rm Int}_{\nu} Z^\dagger_{\nu,-}Z_{\nu,-},
\end{align}
represents the negative-energy part
, and
$H_+$ is the chiral counterpart of $H_-$ corresponding to the positive-energy part.
Note that the degenerate zero-energy modes do not appear in the Hamiltonian.
We remark that 
we do not specify a clear criterion for classifying $X$, $Y$, $Z$,
since changes among them are crossover-like, rather than sharp deformations.
The classification nevertheless gives a useful insight for understanding the physical properties of this model, as we shall argue in the following. 

We next elucidate 
whether the above three types of states, 
$X$, $Y$, and $Z$ are 
localized or not. 
To this end, we investigate the scaling behavior of the inversion participation ratio (IPR), defined as
\begin{align}
P_\nu = \sum_{i} |\psi^\nu (i)|^4.
\end{align}
In Fig.~\ref{fig:ipr_0_pi}(a), 
we plot the IPR for the negative energy sector for $L=200$.
Clearly, the compact right-edge mode has the largest IPR.
The PFLs ($Y$) exhibit the secondary peak.
The remaining states have small IPR, but the BE states ($X$) have slightly larger IPR than 
the rest of the states ($Z$).
In Fig.~\ref{fig:ipr_0_pi}(b), we plot the system size dependence of the $X$, $Y$, and $Z$ states.
For all states, the IPR is fitted by the power function,
$P_\nu = x L^{-y}$.
Again as expected, $Z$ state has the largest exponent ($y = 0.865$), 
$X$ state has the second largest ($y =0.496 $), 
and $Y$ state exhibits the almost localized tendency ($y = 0.220$).
From these results, it is quantitatively clear that the PFL states 
have a distinctively localized character compared with the remaining states.

\begin{figure}[tb]
\begin{center}
\includegraphics[clip,width = 0.95\linewidth]{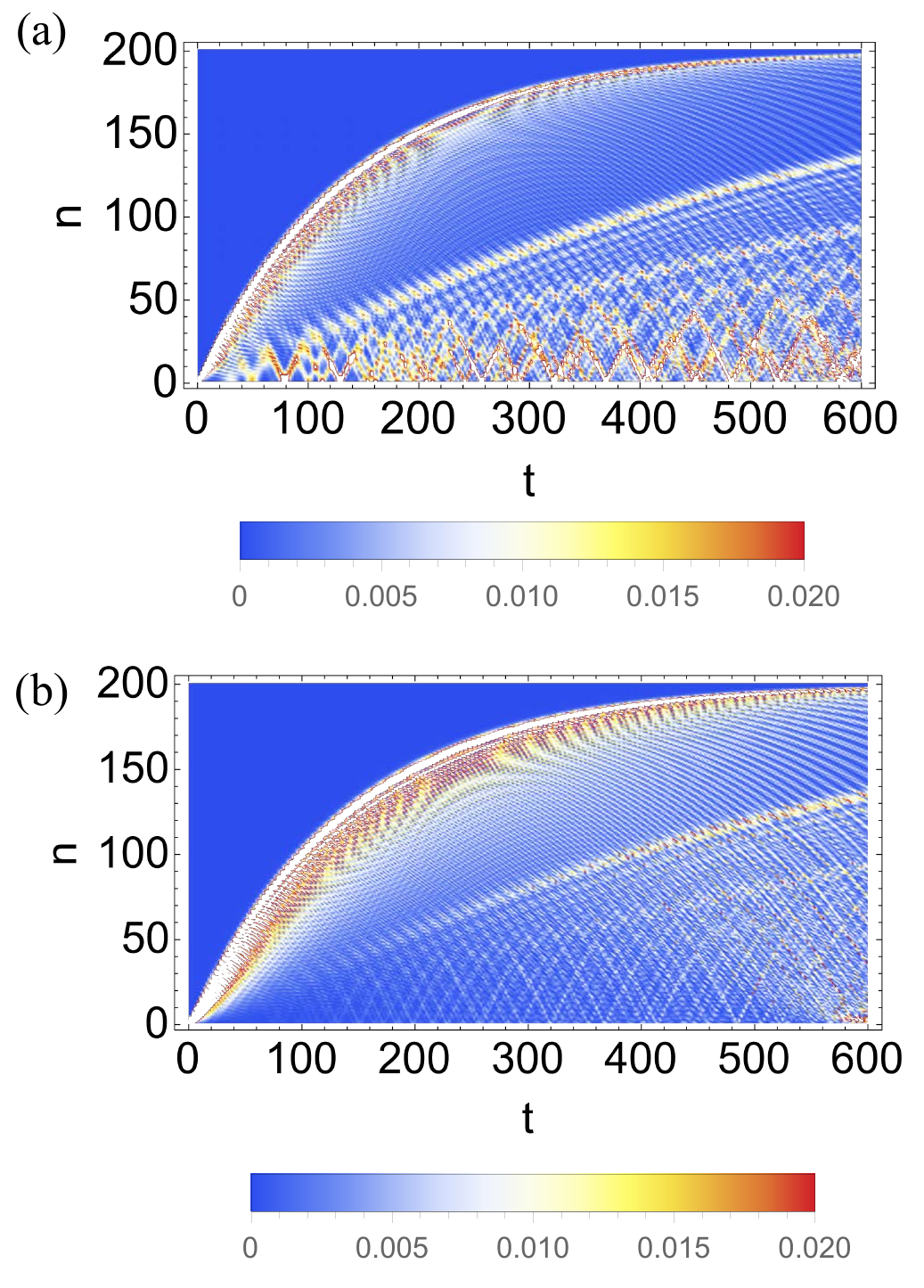}
\vspace{-10pt}
\caption{Time evolution of the probability density distribution per unit cell
for the initial condition, (a) $\phi_{1,\mathrm{A}}(0) = 1$ and 
(b) $\phi_{1,\mathrm{B}}(0)=\phi_{1,\mathrm{C}} = 1/\sqrt{2}$.
}
  \label{fig:dynamics}
 \end{center}
 \vspace{-10pt}
\end{figure}
\begin{figure}[tb]
\begin{center}
\includegraphics[clip,width = 0.9\linewidth]{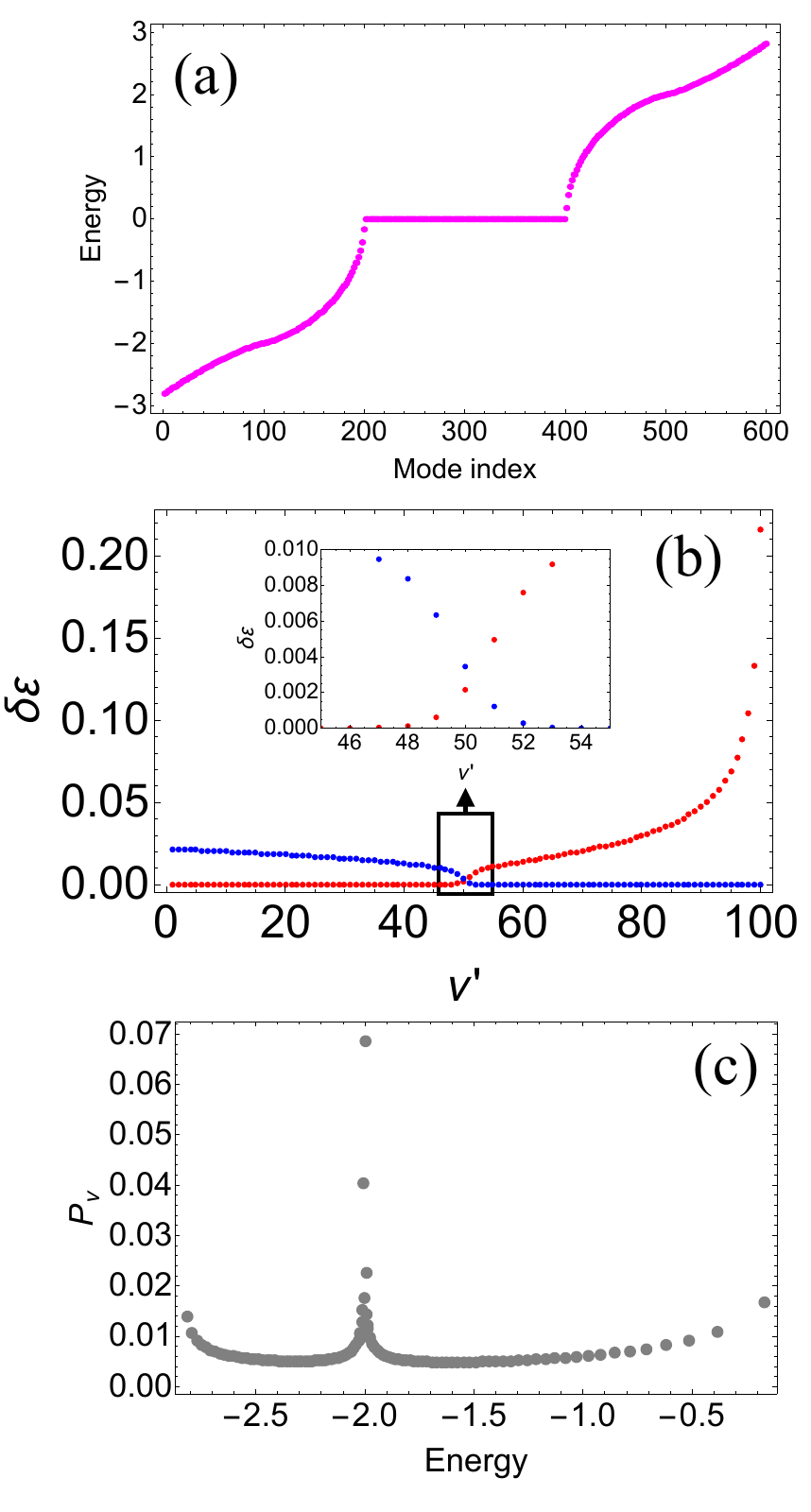}
\vspace{-10pt}
\caption{
Results for $L = 200$ and $(\Phi_{\rm min}, \Phi_{\rm max}) = (0, 2 \pi)$.
(a) Energy spectrum. 
(b) The energy gap between the neighboring eigenenergies (see the main text for its definition). Red [blue] dots represent $\delta \varepsilon^{(1)}_{\nu^\prime}$
[$\delta \varepsilon^{(2)}_{\nu^\prime}$].
(c) IPR for the negative energy sector.
}
  \label{fig:0_2pi}
 \end{center}
 \vspace{-10pt}
\end{figure}
\begin{figure*}[tb]
\begin{center}
\includegraphics[clip,width = 0.9\linewidth]{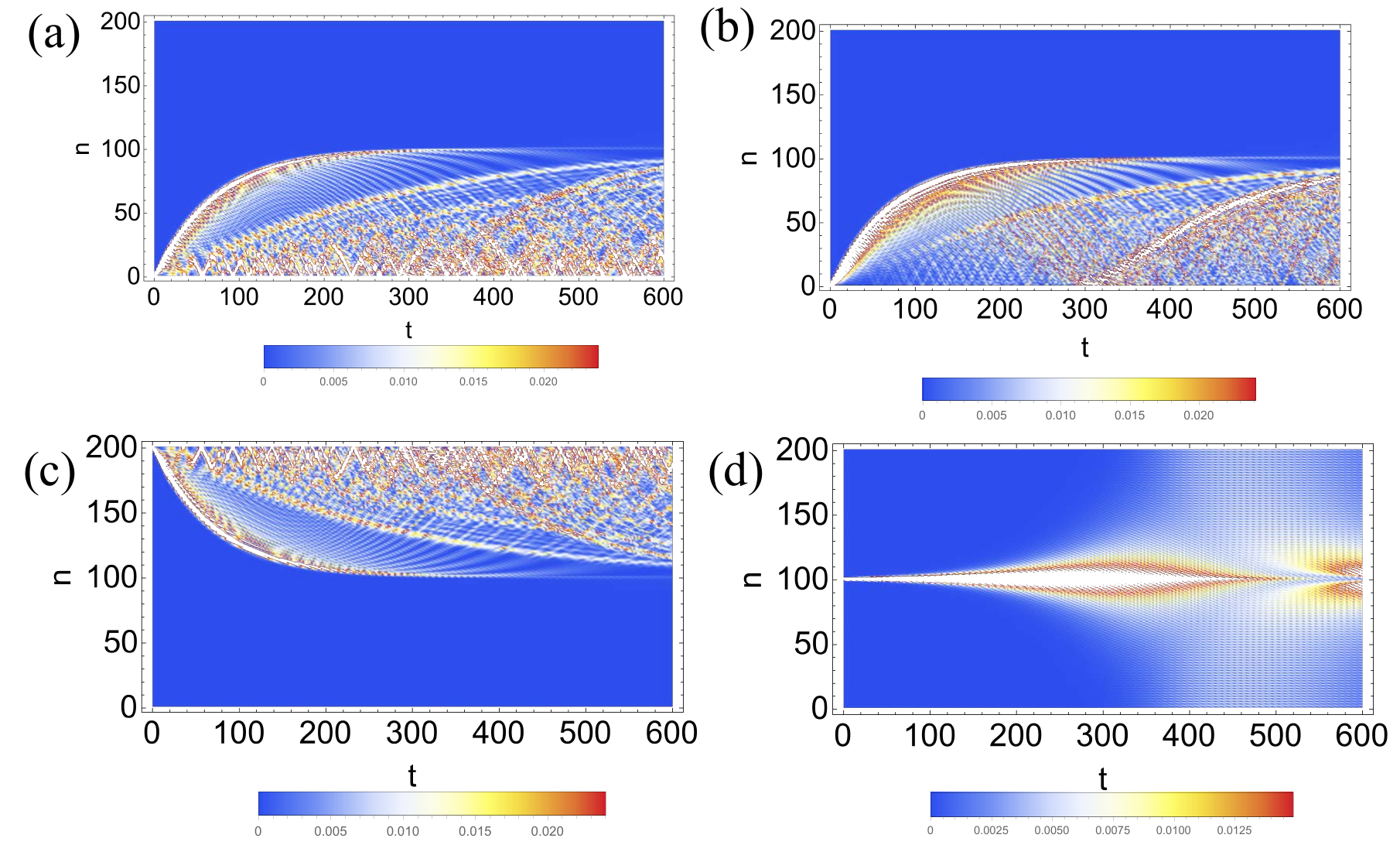}
\vspace{-10pt}
\caption{
Time evolution of the probability density distribution per unit cell
for the initial condition, (a) $\phi_{1,\mathrm{A}}(0) = 1$,
(b) $\phi_{1,\mathrm{B}}(0)=\phi_{1,\mathrm{C}} = 1/\sqrt{2}$,
(c) $\phi_{L+1,\mathrm{A}}(0) = 1$,
and (d) $\phi_{L/2+1,\mathrm{A}}(0) = 1$.
We set $L = 200$ and $(\Phi_{\rm min}, \Phi_{\rm max}) = (0, 2 \pi)$.
}
  \label{fig:0_2pi_d}
 \end{center}
 \vspace{-10pt}
\end{figure*}
\begin{figure*}[tb]
\begin{center}
\includegraphics[clip,width = 1\linewidth]{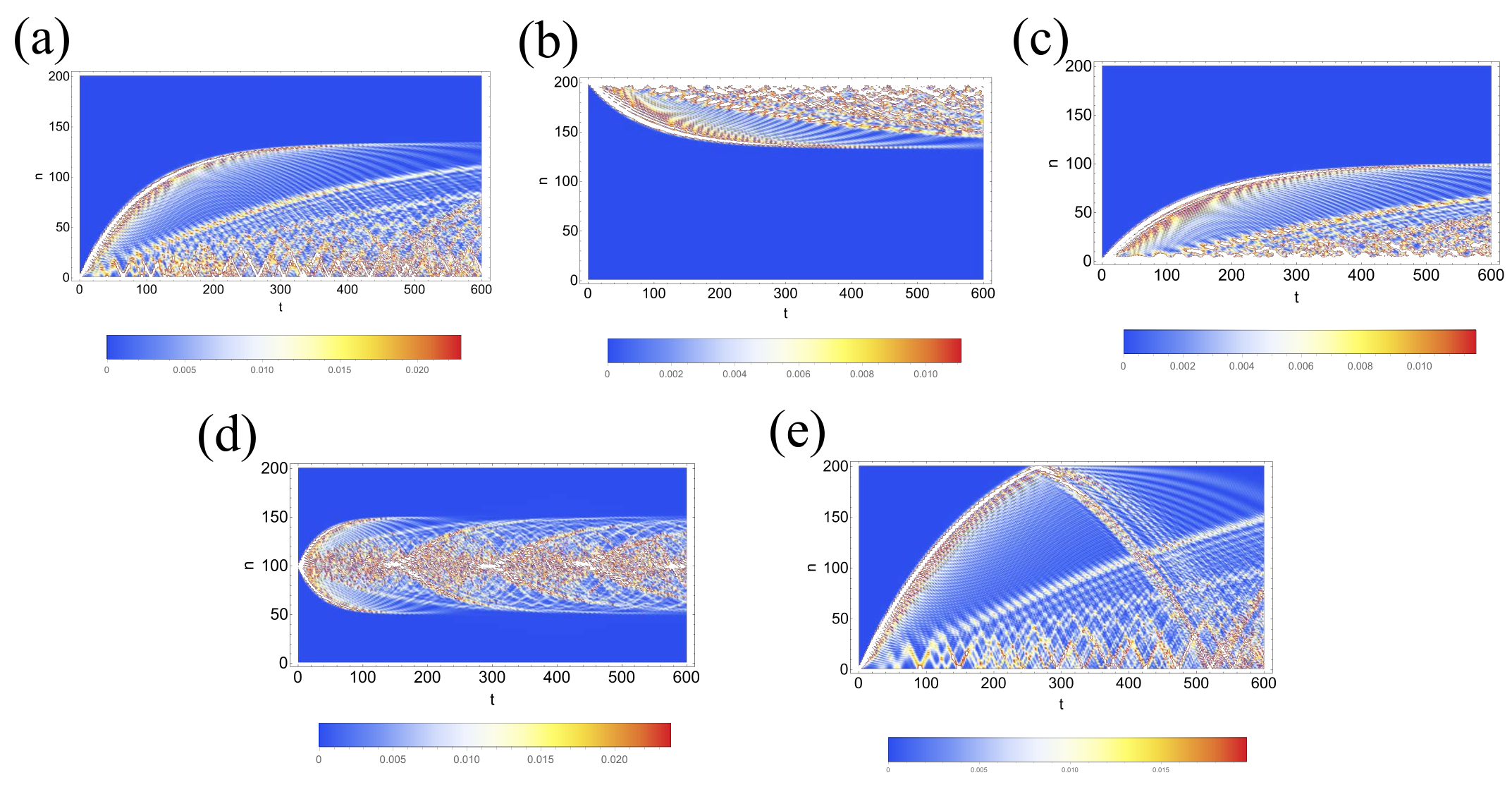}
\vspace{-10pt}
\caption{
Time evolution of the probability density distribution per unit cell
for
(a) $(\Phi_{\rm min},\Phi_{\rm max}) = (0,1.5\pi)$, $\phi_{1,\mathrm{A}}(0) = 1$, 
(b) $(\Phi_{\rm min},\Phi_{\rm max}) = (0,1.5\pi)$, $\phi_{L+1,\mathrm{A}}(0) = 1$, 
(c) $(\Phi_{\rm min},\Phi_{\rm max}) = (0.5\pi,1.5\pi)$, $\phi_{1,\mathrm{A}}(0) = 1$, 
(d) $(\Phi_{\rm min},\Phi_{\rm max}) = (0,4\pi)$, $\phi_{L/2,\mathrm{A}}(0) = 1$,
and (e) $(\Phi_{\rm min},\Phi_{\rm max}) = (0,0.7\pi)$, $\phi_{1,\mathrm{A}}(0) = 1$.
The system size is $L=200$. 
}
  \label{fig:dynamics_various}
 \end{center}
 \vspace{-10pt}
\end{figure*}
Finally, we investigate the single-particle dynamics described by the unitary time evolution.
Let $\ket{\phi(0)}$ be an initial state.
Then, at the time $t$, the state is given as
\begin{align}
\ket{\phi(t)} =  e^{-iH t}\ket{\phi(0)}
=  \sum_{\nu} \phi_{\nu}(t) \ket{\nu}, \label{eq:quench_dynamics_1}
\end{align}
\begin{align}
    \phi_{\nu}(t) = \langle \nu \ket{\phi(t)} = \phi_{\nu}(0) e^{-i\varepsilon_\nu t},\label{eq:quench_dynamics_2}
\end{align}
where $\ket{\nu}:= \alpha^\dagger_\nu \ket{0}$ 
($\ket{0}$ represents the vacuum) and we set $\hbar =1$.

In Fig.~\ref{fig:dynamics}, we plot 
the time evolution of the probability density
per unit cell,
\begin{align}
 N_n(t) = 
 \begin{cases}
 |\phi_{\rm{A},n}(t)|^2 +  |\phi_{\rm{B},n}(t)|^2 + |\phi_{\rm{C},n}(t)|^2 &  1\leq n \leq L \\
  |\phi_{\rm{A},n}(t)|^2 &  n = L+1 \\
  \end{cases},
\end{align}
with $\phi_{i}(t) = \langle i \ket{\phi(t)}  = \bra{0} c_i \ket{\phi(t)}$.
For Fig.~\ref{fig:dynamics}(a) [\ref{fig:dynamics}(b)], 
we set the initial state as $\phi_{1,\mathrm{A}}(0) = 1$ [$\phi_{1,\mathrm{B}}(0)=\phi_{1,\mathrm{C}} = 1/\sqrt{2}$] 
and $0$ otherwise. 
In other words, we set the initial states 
where the particle localizes at the left edge, 
and see how the particle spreads in time evolution.

For both Figs.~\ref{fig:dynamics}(a) and \ref{fig:dynamics}(b), 
we see a characteristic feature of the wavefront.
Namely, for small $t$ ($\lesssim 100$), 
the wavefront moves to the right.
However, on approaching the right edge ($n \geq 120$), 
it slows down and does not reach the right edge even after a very long time (to $t = 600$).
This behavior is understood by the structure of the eigenstates.  
Namely, as indicated in 
Eqs.~(\ref{eq:quench_dynamics_1}) and (\ref{eq:quench_dynamics_2}),
the overlap between the initial state 
and the eigenstates, $\phi_\nu(0)$,
plays a decisive role in the quench dynamics.
In the present case, the initial state has 
a very tiny overlap with the right-edge state and the PFL states.
Since no other eigenstate 
than the right-edge state and the PFL states 
has a large weight near the right edge, 
the particle does not reach the right edge.

It is also worth noting that the degenerate zero modes do not affect the dynamics for Fig.~\ref{fig:dynamics}(a), 
because these zero modes do not have an amplitude at A sites, meaning that the zero modes
have zero overlap to the initial state. 
Considering the fact that the characteristic slow-down of the wavefront is seen in both 
Figs.~\ref{fig:dynamics}(a) and \ref{fig:dynamics}(b), we can conclude that this behavior 
does not originate from the degenerate zero modes.

\subsection{Case II: $(\Phi_{\rm min}, \Phi_{\rm max}) = (0, 2\pi)$}
As another representative case, 
we study the case of $(\Phi_{\rm min}, \Phi_{\rm max}) = (0, 2\pi)$.
It is worth noting that the flux distribution in this case 
satisfies
$\Phi_n \equiv -\Phi_{L+1-n}$ (mod $2\pi$) for $ 1\leq n \leq L/2$,
which means that the right half of the system is the time-reversal 
counterpart of the left half~\cite{remark}.

In Fig.~\ref{fig:0_2pi}(a), 
the energy spectrum for $L= 200$ is plotted.
Remarkably, the energy spectrum looks quite similar to that of Fig.~\ref{fig:DC_model_eigen}(a). 
However, there is a sharp difference from the previous case, that is, 
most of the finite energy modes have a (quasi-)two-fold degeneracy.
To see this, focusing on the negative-energy modes, 
we plot $\delta \varepsilon^{(1)}_{\nu^\prime}:= \varepsilon_{2\nu^\prime}- \varepsilon_{2\nu^\prime-1}$
and $\delta \varepsilon^{(2)}_{\nu^\prime}:= \varepsilon_{2\nu^\prime+1}- \varepsilon_{2\nu^\prime}$ ($\nu^\prime = 1, \cdots, L/2$).
We see that $\delta \varepsilon^{(1)}_{\nu^\prime}$ [$\delta \varepsilon^{(2)}_{\nu^\prime}$] is almost zero for $\nu^\prime \lesssim 50$
[$\nu^\prime \gtrsim 50$]. 
More precisely, setting the numerical threshold
as $\eta = 10^{-5}$, we obtain 
$\delta \varepsilon^{(1)}_{\nu^\prime} < \eta$
for $\nu^\prime \leq 46$ and 
$\delta \varepsilon^{(2)}_{\nu^\prime} < \eta$
for $\nu^\prime \geq  54$.
Hence, as shown in the inset of Fig.~\ref{fig:0_2pi}(b), among the negative energy modes ($\nu = 1, \cdots, 201$),
those with $\nu = 93, \cdots, 107$ are non-degenerate 
while the other modes 
have a two-fold degeneracy within the numerical accuracy. 

In Fig.~\ref{fig:0_2pi}(c), 
we plot the IPR for the negative energy sector.
Note that for a two-fold degenerate pair $\nu_1$ and $\nu_2$,
the IPR is defined as 
\begin{align}
P_{(\nu_1,\nu_2)} = \sum_i \left( \frac{|\psi^{\nu_1}(i)|^2 + |\psi^{\nu_2}(i)|^2}{2} \right)^2.
\end{align}
We again see a sharp peak near $E= -2$, 
which originates from the PFL states localized near $n= L/2$ 
(i.e., the plaquettes with $\Phi_n \sim \pi$).
We note that the edge state is absent in this configuration, 
hence the peak of the IPR at $E=-\sqrt{2}$ seen in Fig.~\ref{fig:ipr_0_pi}(a) is absent. 

In Figs.~\ref{fig:0_2pi_d}(a) and \ref{fig:0_2pi_d}(b), 
we plot the time evolution of the probability density
per unit cell for the same initial state as that 
for Fig.~\ref{fig:dynamics}(a) and 
\ref{fig:dynamics}(b), respectively.
Remarkably, for both cases, 
the particle starting from the left edge 
slows down as it approaches to the center of the system, and it does not reach the right half of the system.

We additionally consider the following two choices of the initial state.
The first one is the case where the particle starts from the right edge [Fig.~\ref{fig:0_2pi_d}(c)].
We see that the particle does not reach the left half of the system.
These behaviors indicate that the PFL states obstruct
the spreading of the particles,
as is the case of $(\Phi_{\rm min}, \Phi_{\rm max}) = (0, \pi)$.
The second one is the case where the particle starts from the center of the system corresponding to the potion of the the nearly-$\pi$-flux plaquette [Fig.~\ref{fig:0_2pi_d}(d)].
We see that the spreading of the particle is highly suppressed for a long time, which is another evidence that the nearly-$\pi$-flux plaquette obstructs the particle dynamics. 

\subsection{Blocked dynamics due to $\pi$-flux plaquette}
From the results of Figs.~\ref{fig:0_2pi}(d)-\ref{fig:0_2pi}(f),
we see that the PFL states
serve as a blockade over which the particle cannot spread. 
To further demonstrate this feature, we plot the time evolution of the probability density for the several combinations 
of $(\Phi_{\rm min}, \Phi_{\rm max})$ and the choices of the initial states. 

In Figs.~\ref{fig:dynamics_various}(a) and \ref{fig:dynamics_various}(b), 
we show the results for 
$(\Phi_{\rm min},\Phi_{\rm max}) = (0,1.5\pi)$,
where the nearly $\pi$-flux plaquette is located at $n \sim (2/3) L $ (i.e., $n\sim 167$ for $L= 200$).
As expected, the particle starting from the left [right] edge does not go
across the opposite side separated by the $\pi$-flux plaquette, as shown in 
Fig.~\ref{fig:dynamics_various}(a) [\ref{fig:dynamics_various}(b)]. 
In Fig.~\ref{fig:dynamics_various}(c), we consider the case where $\Phi_{\rm min} \neq 0$. 
Note that the spectrum around the zero energy modes is gapped, in contrast 
to the cases of $\Phi_{\rm min} = 0$. 
Clearly, the particle is again blocked 
by the $\pi$-flux plaquette, indicating that the blocking of the $\pi$-flux plaquette occurs regardless of the existence of the gap in the energy spectrum. 

In Fig.~\ref{fig:dynamics_various}(d),
we consider the case of $(\Phi_{\rm min},\Phi_{\rm max}) = (0,4\pi)$ 
where there are two
the nearly-$\pi$-flux plaquettes
at $n \sim L/4, (3/4)L$. 
We see that the particle starting at the middle of 
the two nearly-$\pi$-flux plaquettes is confined in the region between them.

Finally, in Fig.~\ref{fig:dynamics_various}(e), 
we consider the case of $(\Phi_{\rm min},\Phi_{\rm max}) = (0,0.7\pi)$ 
that does not contain 
the $\pi$-flux plaquette, 
to clarify the essential role of the $\pi$-flux plaquette.
We see that the blocked dynamics are not seen, 
namely, the wavefront reaches the right edge around $t \sim 300$,
though its velocity slightly decreases as it approaches the right edge.
\begin{figure}[tb]
\begin{center}
\includegraphics[clip,width = 0.95\linewidth]{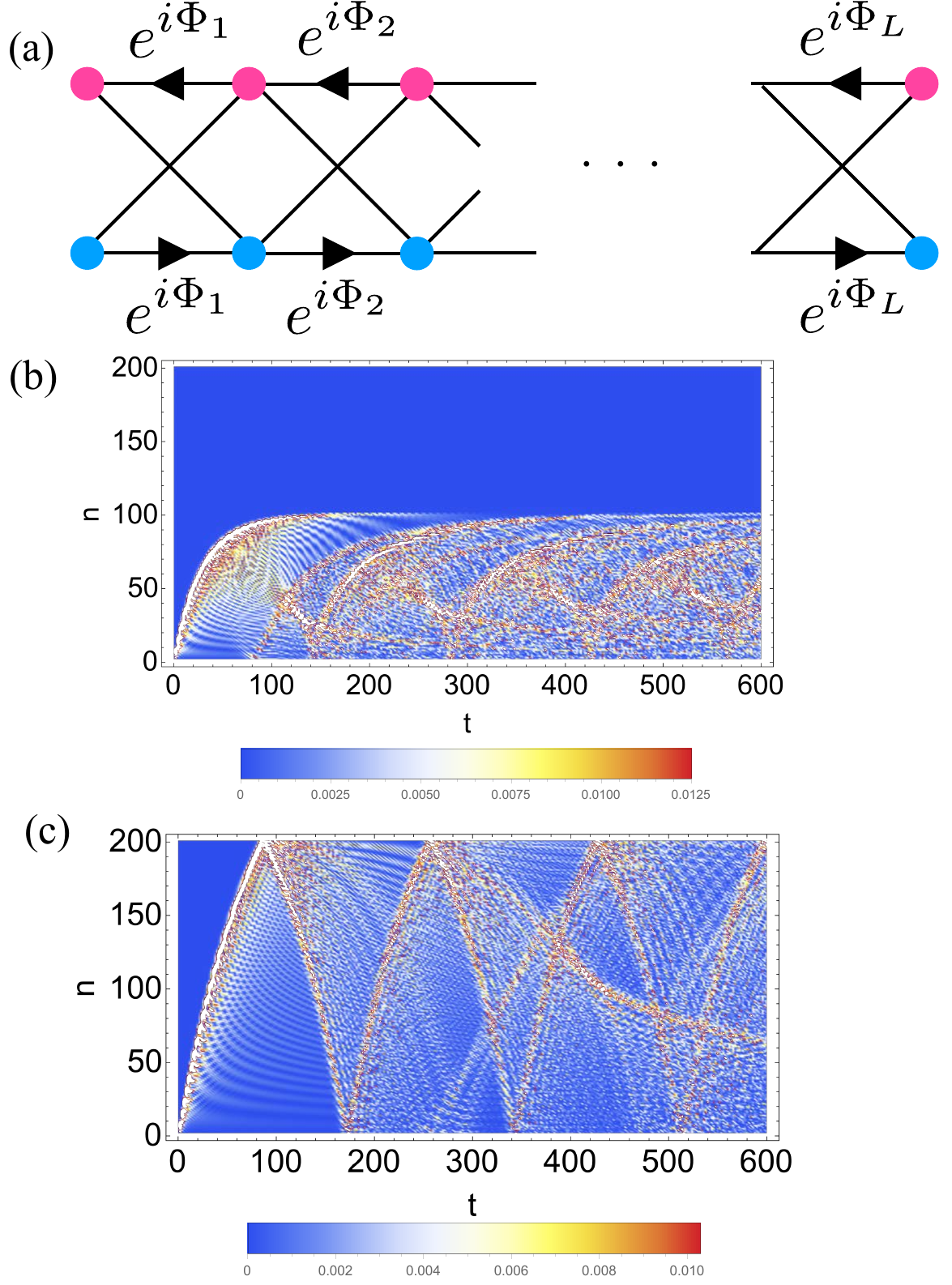}
\vspace{-10pt}
\caption{(a) Schematic figure of the Creutz ladder with spatially increasing phase factors.
Time evolution of the probability density distribution per unit cell
for
(b) $(\Phi_{\rm min},\Phi_{\rm max}) = (0,\pi)$, $\phi_{n=1,\mathrm{u}}(0) = 1$, 
and
(c) $(\Phi_{\rm min},\Phi_{\rm max}) = (0,0.3\pi)$, $\phi_{n=1,\mathrm{u}}(0) = 1$.
Note that $n$ labels the column and the subscript 
``u" stands for the upper row.
We set $L=200$ [the number of sites is $2(L+1) = 402$].}
  \label{fig:CLadder}
 \end{center}
 \vspace{-10pt}
\end{figure}

From these results, the blocked dynamics due to the nearly-$\pi$-flux plaquattes
and the PFL states around them are established. 

\section{Comparison to other models \label{sec:comaprison}}
In this section, we show the results for two additional models 
with spatially-increasing flux, to make a comparison with the diamond chain model. 

\subsection{Creutz ladder}
The Creutz ladder 
has a similar feature to the diamond chain model, 
in that all bands (i.e., two bands in this case) 
become flat at
a specific value of complex hopping~\cite{Creutz1999,Creutz2001,Kuno2020_C,Kuno2020}. 
Here we consider a generalization of the Creutz ladder 
where the phase factors of the complex hoppings 
are spatially increasing as described in Fig.~\ref{fig:CLadder}(a). 
Note that all bands become flat at $\Phi = \frac{\pi}{2}$ for the uniform case.

In Fig.~\ref{fig:CLadder}(b),
we show the particle dynamics for $(\Phi_{\rm min},\Phi_{\rm max} = 0, \pi)$.
Clearly, the particle slows down as approaching $n \sim L/2$, where the value of 
the phase factor is close to that of the all-bands-flat case, exhibiting the similarity to the diamond chain.
For comparison, we also consider the case of $(\Phi_{\rm min},\Phi_{\rm max} = 0, 0.3\pi)$ [Fig.~\ref{fig:CLadder}(c)], 
where none of the phase factors corresponds to the all-bands-flat case.
In this case, the blocking of the particle does not occur and it reaches the right edge, 
which also resembles the result of the diamond chain.

\subsection{Two-leg ladder}
\begin{figure*}[tb]
\begin{center}
\includegraphics[clip,width = 0.95\linewidth]{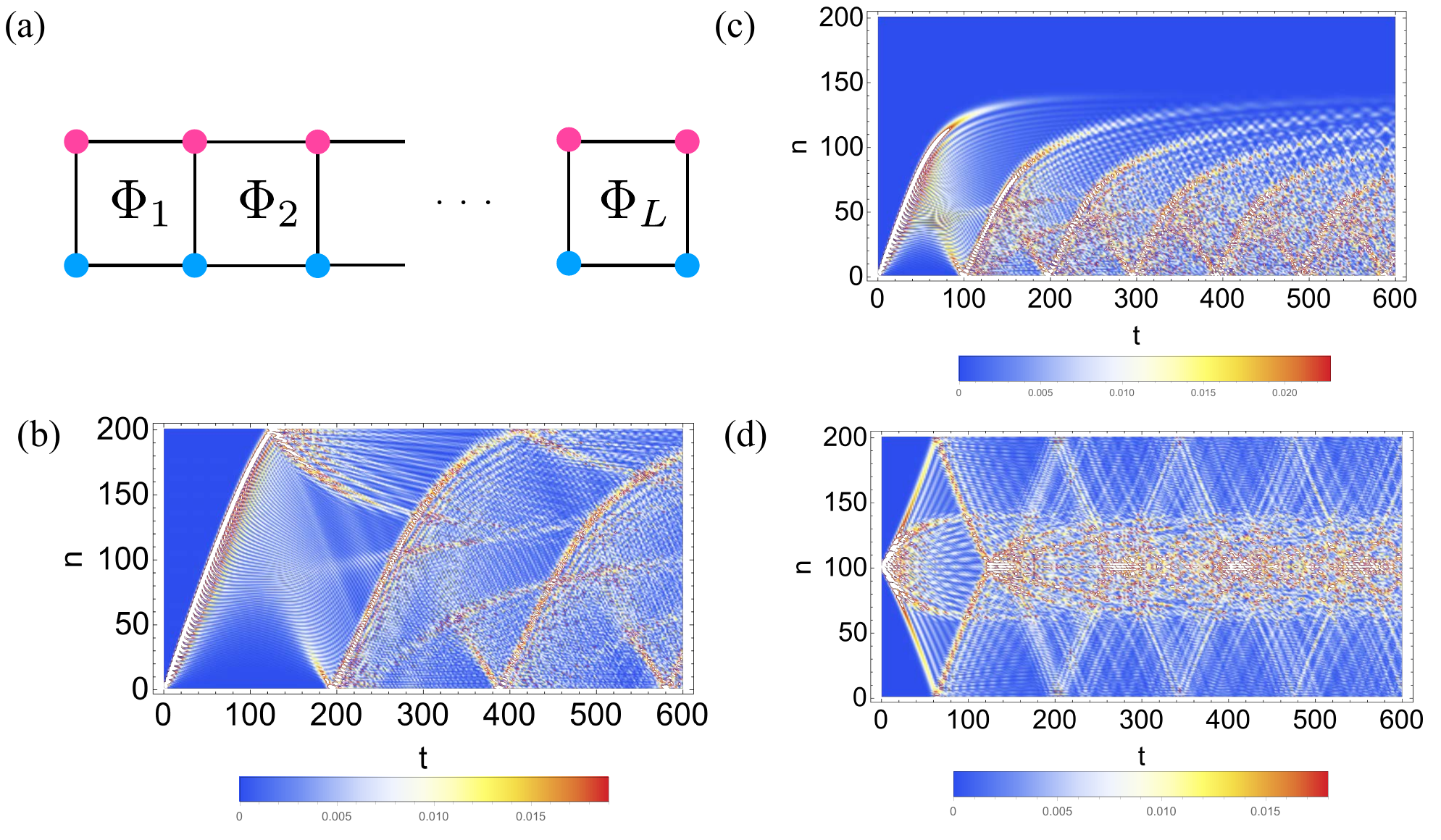}
\vspace{-10pt}
\caption{(a) Schematic figure of the two-leg ladder with spatially increasing flux. Time evolution of the probability density distribution per unit cell for
(b) $(\Phi_{\rm min},\Phi_{\rm max}) = (0,\pi)$, $\phi_{n=1,\mathrm{u}}(0) = 1$
(c) $(\Phi_{\rm min},\Phi_{\rm max}) = (0,2\pi)$, $\phi_{n=1,\mathrm{u}}(0) = 1$
and (d) $(\Phi_{\rm min},\Phi_{\rm max}) = (0,2\pi)$, $\phi_{n=L/2+1,\mathrm{u}}(0) = 1$.
Note that $n$ labels the column and the subscript ``u" (``l") stands for the upper (lower) row. We set $L=200$ [the number of sites is $2(L+1) = 402$].
}
  \label{fig:TLLadder}
 \end{center}
 \vspace{-10pt}
\end{figure*}
We next consider the two-leg ladder model with a magnetic flux [Fig.~\ref{fig:TLLadder}(a)].
In contrast to the diamond chain and Creutz lattice, in the uniform case, 
the complete flat bands do not appear at any value of the flux. 

In Figs.~\ref{fig:TLLadder}(b) and \ref{fig:TLLadder}(c), 
we show the particle dynamics starting from the left edge,
for $(\Phi_{\rm min},\Phi_{\rm max}) = (0,\pi)$ and $(\Phi_{\rm min},\Phi_{\rm max}) = (0,2\pi)$, 
respectively. 
In Fig.~\ref{fig:TLLadder}(b), 
the particle exhibits the standard spreading dynamics, reaches the right edge after a certain time,
and the reflected wave arises, as expected. 
Meanwhile, in Fig.~\ref{fig:TLLadder}(c), 
the particle sharply slows down around $n \sim 130$,
similar to the diamond chain and the Creutz ladder.
This behavior is non-trivial because 
the sharply localized eigenstates 
are expected to arise for any values of flux.
For further comparison, we show the particle dynamics with the initial position being the middle of the system 
in Fig.~\ref{fig:TLLadder}(d).
We see that the blocking of the particle dynamics is much weaker than the diamond chain case, and 
not a few amounts of the particle density propagate to the left and right edges.
Combining these results, 
we speculate that there can be a universal (i.e., lattice independent) mechanism of the blocking dynamics by the spatially increasing flux even without the localized wave functions unique to the all-flat-band systems,
but the degree of blocking is not as strong 
as that of the all-flat-band systems such as the diamond chain and the Creutz ladder.
Further studies on 
a possible mechanism are necessary to
extract the uniqueness of the all-bands-flat lattices.  

\subsection{Early time dynamics \label{sec:earlytime}}
We further make a comparison among three models from a different point of view.
Namely, we focus on early time dynamics for each model. 
Study of particle or correlation spreading is interesting since real experiments can capture such a spreading. 
For example, a recent optical lattice experiment \cite{Cheneau2012} 
observed the spread of correlation between doublon and holon in a quench dynamics and found a linear-like propagation of it. Also a cloud spreading has been investigated in detail \cite{Ronzheimer2013}. Some theoretical works about quench dynamics of particle spreading in early time has been reported \cite{Jreissaty2013,Vidmar2013}. 
Motivated by these works, we focus on the early time dynamics 
of the flat-band and the non-flat-band models. 
Several behaviors are observed from our numerical results:
\begin{enumerate}
    \item In the flat band model with linear increase flux $(\Phi_{\rm min},\Phi_{\rm max}) = (0,\pi)$ [as shown in Fig.5 (a)], its dynamics exhibits no-linear right cone spreading. We expect that even if an initial particle is put on any position, such a dynamics occurs. 
    \item
    In the two-leg ladder model with linear increase $(\Phi_{\rm min},\Phi_{\rm max}) = (0,\pi)$, 
    linear-like spreading in $0\leq t \leq 150$ is observed even if an initial particle is put on any position [as shown in Fig.~\ref{fig:TLLadder}(b)]. 
    \item
    In the two-leg ladder model with linear increase $(\Phi_{\rm min},\Phi_{\rm max}) = (0,2\pi)$, 
    when the initial particle is set around the $\pi$-flux, the particle matter wave clearly exhibits linear wave-front 
    [Fig.~\ref{fig:TLLadder}(d)]. This is significantly different from that of the flat band case, where the initial particle is not spread, 
    highly bounded around $\pi$-flux as shown in Fig.~\ref{fig:0_2pi_d}(d).
\end{enumerate}
In particular, the observation 3 in the above implies a highly localized eigenstate exists 
around $\pi$-flux in flat-band model while such a state does not appear in the two-leg ladder model. 
Such a highly localized state 
gives significant effects to the dynamics with particle initially set around $\pi$-flux. 

These behaviors, in particular, the difference between the flat-band model and conventional dispersive band models, 
can be observed in a real experiment such as photonic waveguides.

\section{Summary} \label{sec:summary}
We have investigated the characteristic structures of the eigenstates
and resulting dynamics in the diamond chain model 
with spatially increasing flux.
For the uniform flux case, the remarkable feature of the diamond chain model
is the realization of the all-band-flat system at $\pi$-flux.
This feature is succeeded to the spatially-increasing-flux case,
in that the sharply localized eigenstates emerge around the $\pi$-flux plaquette.
Consequently, the $\pi$-flux plaquette serves as a blockade of the particle dynamics.
Indeed, by investigating the particle dynamics with the localized eigenstates, 
we find that the particle slows down as approaching the $\pi$-flux plaquette.
This behavior of the partilce dynamics is unique to the present models, 
which does not resemble 
any one of 
the conventional spreading dynamics for itinerant systems, 
the Bloch oscillation for the Wannier-Stark-type localized systems,
or the complete localization for the Aharonov-Bohm cages.  

We close this paper by addressing future directions of research.
As for the single-particle dynamics, various patterns of spatially-varying flux, such as a random flux or quasi-periodic flux, 
will be sources of unconventional features, 
which we think are worth being studied.  
To investigate eigenstate properties of the squared Hamiltonian gives an insight to understand the localization properties of the eigenstates of the original model.
Considering the many-particle system under the present setup is another interesting direction because the $\pi$-flux blockade serves as a novel mechanism of confining a particle which 
will lead to slow thermalization or disorder-free localization.
Finally, the experimental realization 
of the present model will also be an important issue.
The photonic waveguides and ultracold atoms~\cite{Bloch2008,Aidelsburger2011} will be possible platforms due to the tunability of the effective magnetic flux.
Quite recently, the electric circuit realization of the $\pi$-flux diamond chain was also reported~\cite{Zhang2023}, which may offer another platform of the experimental realization of our model.

\acknowledgments
This work is supported by JST CREST Grant Number JPMJCR19T1, and by JSPS KAKENHI Grant Numbers 
23H01091 (Y. H.) and 23K13026 (Y.K.).

\appendix
\section{Bulk spectrum of diamond chain with flux \label{app:bulkband}}
\begin{figure}[tb]
\begin{center}
\includegraphics[clip,width = 0.95\linewidth]{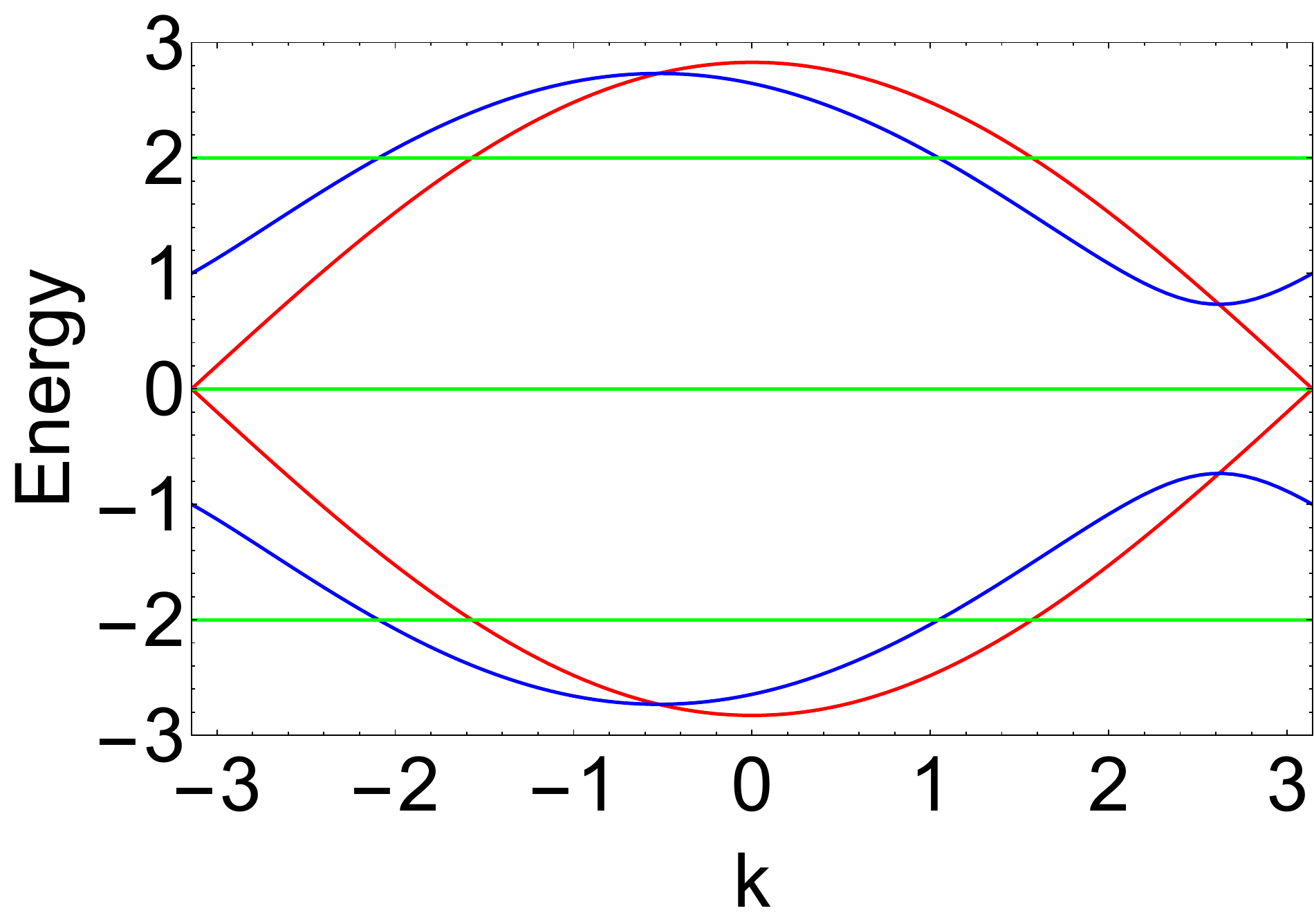}
\vspace{-10pt}
\caption{Band structure for the diamond chain model with uniform flux. 
The red, blue, and green lines are for $\Phi = 0$, $\pi/3$, and $\pi$, respectively.}
  \label{fig:DC_band}
 \end{center}
 \vspace{-10pt}
\end{figure}
In this appendix, 
we review the bulk spectrum of the diamond chain 
in the presence of the uniform flux $\Phi$.
For the uniform flux, the transitional invariance
is preserved, hence we use the momentum space description. 
The Bloch Hamiltonian reads
\begin{align}
H(k) = \begin{pmatrix}
0 & 1 + e^{-i (k+\Phi)} & 1 + e^{-ik} \\
 1 + e^{i (k+\Phi)} & 0 & 0 \\
  1 + e^{ik} & 0 & 0 \\
\end{pmatrix}.
\end{align}
As mentioned in the main text, the model preserves 
the chiral symmetry.
In the momentum-space picture, 
this symmetry can be represented by  
$\bar{g} H(k) \bar{g} = - H(k)$, where
$\bar{g} = \mathrm{diag}(1,-1,-1)$.
Since $|\mathrm{Tr} (\bar{g}) | = 1$, there exists a zero-energy mode for any $k$.

The dispersion relation can be obtained analytically by, again, taking the square of the Hamiltonian:
\begin{align}
   H^2(k)  = \begin{pmatrix}
|f_1(k)|^2 + |f_2(k)|^2 & 0 & 0 \\
0& |f_1(k)|^2  & f_1^\ast(k) f_2(k) \\
 0  & f_1(k) f^\ast_2(k) & |f_2(k)|^2 \\
\end{pmatrix}, \label{eq:ham_square}
\end{align}
where $f_1(k) = 1 + e^{-i (k+\Phi)}$ and 
$f_2(k) = 1 + e^{-i k}$.
We can easily find from Eq.~(\ref{eq:ham_square}) that the eigenenergies of $H^2 (k)$ is $0$ and $|f_1(k)|^2 + |f_2(k)|^2$ (doubly degenerate). Consequently, the eigenenergies and the eigenvectors of $H(k)$ are given as
\begin{subequations}
\begin{align}
E_{\pm} (k)=& \pm \sqrt{|f_1(k)|^2 + |f_2 (k)|^2}, \notag \\
\bm{u}_{k,\pm} =& 
\frac{1}{\sqrt{2(|f_1(k)|^2 + |f_2 (k)|^2)}}
\begin{pmatrix}
 \sqrt{|f_1(k)|^2 + |f_2 (k)|^2} \\
 \pm f^\ast_1(k)  \\
 \pm f^\ast_2 (k)\\
\end{pmatrix}, \label{eq:bloch_1}
\end{align}
\begin{align}
E_{0} (k) =0 ,
\hspace{.5mm} \bm{u}_{k,0} =
\frac{1}{\sqrt{|f_1(k)|^2 + |f_2 (k)|^2}}
\begin{pmatrix}
0 \\
f_2(k) \\
-f_1(k) \\
\end{pmatrix}.\label{eq:bloch_2}
\end{align}
\end{subequations}
It should be noted that at $\Phi = \pi$, we have 
$E_{\pm} (k) = \pm 2 $ for any $k$, which means that
all bands are flat in this case.
In Fig.~\ref{fig:DC_band}, we plot the band structures 
for $\Phi = 0$ (red), $\pi/3$ (blue), and $\pi$ (green).

\begin{figure}[tb]
\begin{center}
\includegraphics[clip,width = 0.95\linewidth]{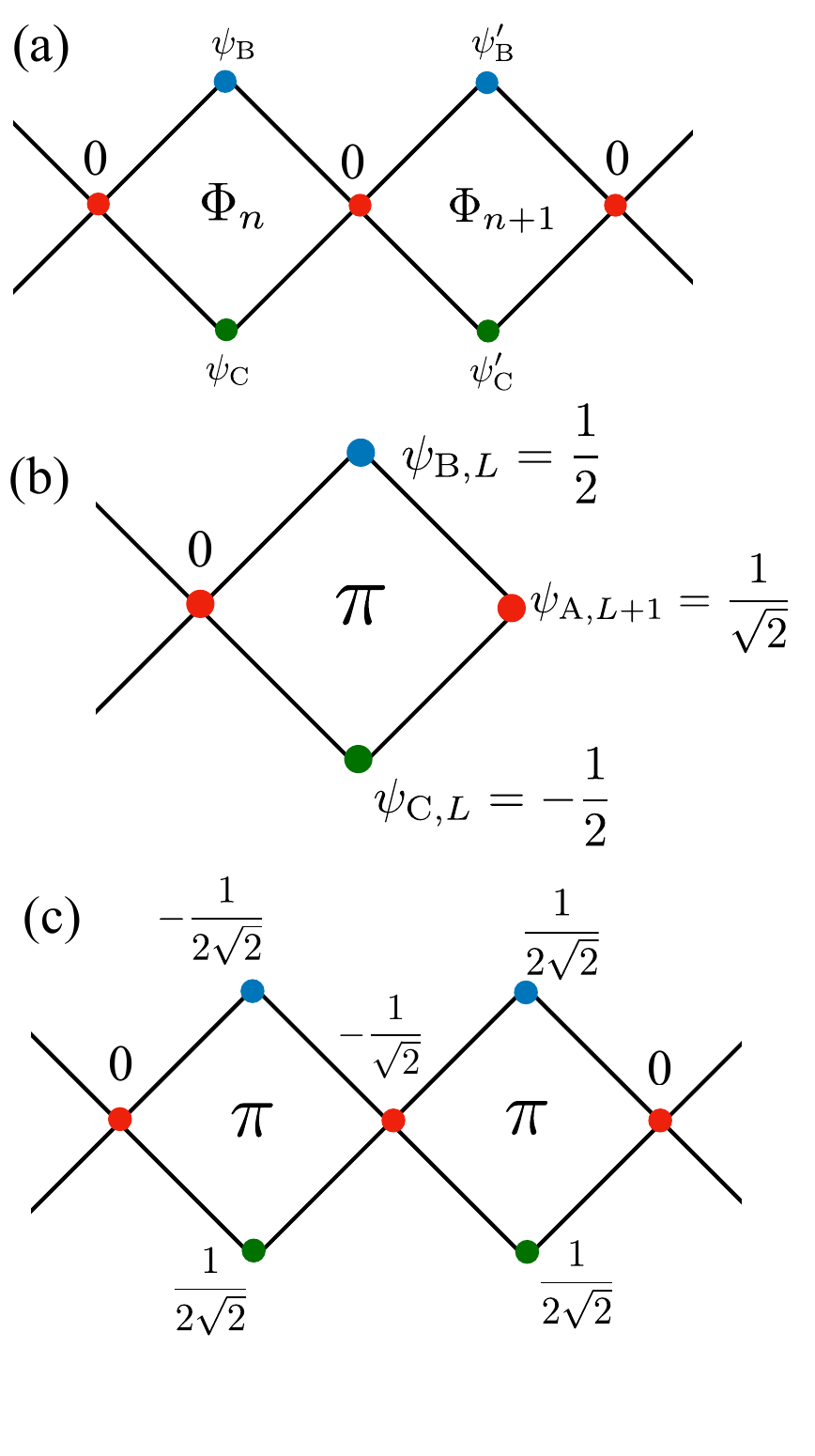}
\vspace{-10pt}
\caption{(a) Schematic figures of (a) the CLS with $E=0$, 
(b) the right-edge state, and
(c) the CLS with $E=-2$ for the uniform $\pi$-flux.}
  \label{fig:cls}
 \end{center}
 \vspace{-10pt}
\end{figure}
\section{CLS at $E=0$ \label{app:CLS}}
In this appendix, we elucidate the CLS at $E=0$. 
The guiding principle of constructing the CLS is to
set the amplitudes at A sites to be zero. 
For the configuration of Fig.~\ref{fig:cls}(a),
we obtain the compact wave function with finite amplitudes on only four sites. 
For its solution, the above assumption 
leads to the following three equations:
\begin{subequations}
\begin{align}
 \psi_{\rm B} +  \psi_{\rm C} =0,
\end{align}
\begin{align}
   e^{-i\Phi_n} \psi_{\rm B} +  \psi_{\rm C} 
    + \psi^\prime_{\rm B} +  \psi^\prime_{\rm C}=0,
\end{align}
and
\begin{align}
    e^{-i\Phi_{n+1}}  \psi^\prime_{\rm B} +  \psi^\prime_{\rm C}=0.
\end{align}
\end{subequations}
From these equations, we obtain the wave function of the CLS:
\begin{align}
    \left( \psi_{\rm B}, \psi_{\rm C},\psi^\prime_{\rm B},\psi^\prime_{\rm C} \right)
     = \frac{1}{\mathcal{N}} \left(1, -1, - \frac{x_1}{x_2}, e^{-i\Phi_{n+1}} \frac{x_1}{x_2} \right), \label{eq:CLS_analytic}
\end{align}
where $\mathcal{N}$ is the normalization factor,
$x_1 = e^{-i\Phi_n}-1$,
and $x_2 =1- e^{-i\Phi_{n+1}} $.
It is worth noting that the CLSs are in general not orthogonal to each other,
since the neighboring CLSs overlap.
We also note that
the solution is not valid
when $\Phi_{n+1} = 0$ because $x_2 = 0$.
In fact, in this case, the CLS is given as
\begin{align}
    \left( \psi_{\rm B}, \psi_{\rm C},\psi^\prime_{\rm B},\psi^\prime_{\rm C} \right)
     = \frac{1}{\sqrt{2}} \left(0,0,1, -1 \right), \label{eq:CLS_analytic_2}
\end{align}
which indicates 
that the CLS has finite amplitude on only two sites rather than four sites. 

\section{Right edge state for $\Phi_{\rm max} = \pi$ \label{app:right-edge}}
Here we remark on the right edge state for $\Phi_{\rm max} = \pi$.
For the uniform $\pi$-flux model, the compact edge states
with the eigenenergy $\pm \sqrt{2}$ appear~\cite{Kremer2020}. 
Due to the compact nature, 
we have the same edge states even in the present case of the increasing flux with $\Phi_{\rm max} = \pi$. 
To be specific, for the configuration of Fig.~\ref{fig:cls}(c),
the right edge state has finite amplitudes 
at only three sites, i.e., $(\rm {B},L)$, $(\rm {C},L)$, and $(\rm {A},{L+1})$.
For the eigenstate with $E = -\sqrt{2}$, the wave function is given as
\begin{align}
\left( \psi_{\rm{A},L+1}, \psi_{\rm{B},L}, \psi_{\rm{C},L} \right) = \left(\frac{1}{\sqrt{2}}, \frac{1}{2},-\frac{1}{2} \right).  
\end{align}

\section{Finite-energy CLS for $\pi$-flux \label{app:finiteCLS}}
As mentioned in Appendix~\ref{app:bulkband}, the case of the uniform flux with $\Phi = \pi$ is special in that all bands are completely flat. 
Therefore, the states with $E = \pm 2$ can also be
given by the set of CLSs. 
In Fig.~\ref{fig:cls}(b), 
we present the wave function for $E=-2$,
which can be obtained by solving the Sch\"{o}dinger equation explicitly.

\bibliographystyle{apsrev4-2}
\bibliography{diamond_chain}

\begin{thebibliography}{65}%
\makeatletter
\providecommand \@ifxundefined [1]{%
 \@ifx{#1\undefined}
}%
\providecommand \@ifnum [1]{%
 \ifnum #1\expandafter \@firstoftwo
 \else \expandafter \@secondoftwo
 \fi
}%
\providecommand \@ifx [1]{%
 \ifx #1\expandafter \@firstoftwo
 \else \expandafter \@secondoftwo
 \fi
}%
\providecommand \natexlab [1]{#1}%
\providecommand \enquote  [1]{``#1''}%
\providecommand \bibnamefont  [1]{#1}%
\providecommand \bibfnamefont [1]{#1}%
\providecommand \citenamefont [1]{#1}%
\providecommand \href@noop [0]{\@secondoftwo}%
\providecommand \href [0]{\begingroup \@sanitize@url \@href}%
\providecommand \@href[1]{\@@startlink{#1}\@@href}%
\providecommand \@@href[1]{\endgroup#1\@@endlink}%
\providecommand \@sanitize@url [0]{\catcode `\\12\catcode `\$12\catcode
  `\&12\catcode `\#12\catcode `\^12\catcode `\_12\catcode `\%12\relax}%
\providecommand \@@startlink[1]{}%
\providecommand \@@endlink[0]{}%
\providecommand \url  [0]{\begingroup\@sanitize@url \@url }%
\providecommand \@url [1]{\endgroup\@href {#1}{\urlprefix }}%
\providecommand \urlprefix  [0]{URL }%
\providecommand \Eprint [0]{\href }%
\providecommand \doibase [0]{https://doi.org/}%
\providecommand \selectlanguage [0]{\@gobble}%
\providecommand \bibinfo  [0]{\@secondoftwo}%
\providecommand \bibfield  [0]{\@secondoftwo}%
\providecommand \translation [1]{[#1]}%
\providecommand \BibitemOpen [0]{}%
\providecommand \bibitemStop [0]{}%
\providecommand \bibitemNoStop [0]{.\EOS\space}%
\providecommand \EOS [0]{\spacefactor3000\relax}%
\providecommand \BibitemShut  [1]{\csname bibitem#1\endcsname}%
\let\auto@bib@innerbib\@empty
\bibitem [{\citenamefont {Hatsugai}\ and\ \citenamefont
  {Sugi}(2001)}]{Hatsugai2001}%
  \BibitemOpen
  \bibfield  {author} {\bibinfo {author} {\bibfnamefont {Y.}~\bibnamefont
  {Hatsugai}}\ and\ \bibinfo {author} {\bibfnamefont {A.}~\bibnamefont
  {Sugi}},\ }\href {https://doi.org/10.1142/S0217979201004885} {\bibfield
  {journal} {\bibinfo  {journal} {International Journal of Modern Physics B}\
  }\textbf {\bibinfo {volume} {15}},\ \bibinfo {pages} {2045} (\bibinfo {year}
  {2001})}\BibitemShut {NoStop}%
\bibitem [{\citenamefont {Mazza}\ \emph {et~al.}(2015)\citenamefont {Mazza},
  \citenamefont {Aidelsburger}, \citenamefont {Tu}, \citenamefont {Goldman},\
  and\ \citenamefont {Burrello}}]{Mazza2015}%
  \BibitemOpen
  \bibfield  {author} {\bibinfo {author} {\bibfnamefont {L.}~\bibnamefont
  {Mazza}}, \bibinfo {author} {\bibfnamefont {M.}~\bibnamefont {Aidelsburger}},
  \bibinfo {author} {\bibfnamefont {H.-H.}\ \bibnamefont {Tu}}, \bibinfo
  {author} {\bibfnamefont {N.}~\bibnamefont {Goldman}},\ and\ \bibinfo {author}
  {\bibfnamefont {M.}~\bibnamefont {Burrello}},\ }\href
  {https://doi.org/10.1088/1367-2630/17/10/105001} {\bibfield  {journal}
  {\bibinfo  {journal} {New Journal of Physics}\ }\textbf {\bibinfo {volume}
  {17}},\ \bibinfo {pages} {105001} (\bibinfo {year} {2015})}\BibitemShut
  {NoStop}%
\bibitem [{\citenamefont {Meier}\ \emph {et~al.}(2016)\citenamefont {Meier},
  \citenamefont {An},\ and\ \citenamefont {Gadway}}]{Meier2016}%
  \BibitemOpen
  \bibfield  {author} {\bibinfo {author} {\bibfnamefont {E.~J.}\ \bibnamefont
  {Meier}}, \bibinfo {author} {\bibfnamefont {F.~A.}\ \bibnamefont {An}},\ and\
  \bibinfo {author} {\bibfnamefont {B.}~\bibnamefont {Gadway}},\ }\href
  {https://doi.org/10.1038/ncomms13986} {\bibfield  {journal} {\bibinfo
  {journal} {Nature Communications}\ }\textbf {\bibinfo {volume} {7}},\
  \bibinfo {pages} {13986} (\bibinfo {year} {2016})}\BibitemShut {NoStop}%
\bibitem [{\citenamefont {Wang}\ \emph {et~al.}(2017)\citenamefont {Wang},
  \citenamefont {Zhang}, \citenamefont {Chen}, \citenamefont {Yu},\ and\
  \citenamefont {Zhai}}]{Wang2017}%
  \BibitemOpen
  \bibfield  {author} {\bibinfo {author} {\bibfnamefont {C.}~\bibnamefont
  {Wang}}, \bibinfo {author} {\bibfnamefont {P.}~\bibnamefont {Zhang}},
  \bibinfo {author} {\bibfnamefont {X.}~\bibnamefont {Chen}}, \bibinfo {author}
  {\bibfnamefont {J.}~\bibnamefont {Yu}},\ and\ \bibinfo {author}
  {\bibfnamefont {H.}~\bibnamefont {Zhai}},\ }\href
  {https://doi.org/10.1103/PhysRevLett.118.185701} {\bibfield  {journal}
  {\bibinfo  {journal} {Phys. Rev. Lett.}\ }\textbf {\bibinfo {volume} {118}},\
  \bibinfo {pages} {185701} (\bibinfo {year} {2017})}\BibitemShut {NoStop}%
\bibitem [{\citenamefont {Cardano}\ \emph {et~al.}(2017)\citenamefont
  {Cardano}, \citenamefont {D'Errico}, \citenamefont {Dauphin}, \citenamefont
  {Maffei}, \citenamefont {Piccirillo}, \citenamefont {de~Lisio}, \citenamefont
  {De~Filippis}, \citenamefont {Cataudella}, \citenamefont {Santamato},
  \citenamefont {Marrucci}, \citenamefont {Lewenstein},\ and\ \citenamefont
  {Massignan}}]{Cardano2017}%
  \BibitemOpen
  \bibfield  {author} {\bibinfo {author} {\bibfnamefont {F.}~\bibnamefont
  {Cardano}}, \bibinfo {author} {\bibfnamefont {A.}~\bibnamefont {D'Errico}},
  \bibinfo {author} {\bibfnamefont {A.}~\bibnamefont {Dauphin}}, \bibinfo
  {author} {\bibfnamefont {M.}~\bibnamefont {Maffei}}, \bibinfo {author}
  {\bibfnamefont {B.}~\bibnamefont {Piccirillo}}, \bibinfo {author}
  {\bibfnamefont {C.}~\bibnamefont {de~Lisio}}, \bibinfo {author}
  {\bibfnamefont {G.}~\bibnamefont {De~Filippis}}, \bibinfo {author}
  {\bibfnamefont {V.}~\bibnamefont {Cataudella}}, \bibinfo {author}
  {\bibfnamefont {E.}~\bibnamefont {Santamato}}, \bibinfo {author}
  {\bibfnamefont {L.}~\bibnamefont {Marrucci}}, \bibinfo {author}
  {\bibfnamefont {M.}~\bibnamefont {Lewenstein}},\ and\ \bibinfo {author}
  {\bibfnamefont {P.}~\bibnamefont {Massignan}},\ }\href
  {https://doi.org/10.1038/ncomms15516} {\bibfield  {journal} {\bibinfo
  {journal} {Nature Communications}\ }\textbf {\bibinfo {volume} {8}},\
  \bibinfo {pages} {15516} (\bibinfo {year} {2017})}\BibitemShut {NoStop}%
\bibitem [{\citenamefont {Gong}\ and\ \citenamefont {Ueda}(2018)}]{Gong2018}%
  \BibitemOpen
  \bibfield  {author} {\bibinfo {author} {\bibfnamefont {Z.}~\bibnamefont
  {Gong}}\ and\ \bibinfo {author} {\bibfnamefont {M.}~\bibnamefont {Ueda}},\
  }\href {https://doi.org/10.1103/PhysRevLett.121.250601} {\bibfield  {journal}
  {\bibinfo  {journal} {Phys. Rev. Lett.}\ }\textbf {\bibinfo {volume} {121}},\
  \bibinfo {pages} {250601} (\bibinfo {year} {2018})}\BibitemShut {NoStop}%
\bibitem [{\citenamefont {Zhang}\ \emph {et~al.}(2019)\citenamefont {Zhang},
  \citenamefont {Zhang},\ and\ \citenamefont {Liu}}]{Zhang2019}%
  \BibitemOpen
  \bibfield  {author} {\bibinfo {author} {\bibfnamefont {L.}~\bibnamefont
  {Zhang}}, \bibinfo {author} {\bibfnamefont {L.}~\bibnamefont {Zhang}},\ and\
  \bibinfo {author} {\bibfnamefont {X.-J.}\ \bibnamefont {Liu}},\ }\href
  {https://doi.org/10.1103/PhysRevA.99.053606} {\bibfield  {journal} {\bibinfo
  {journal} {Phys. Rev. A}\ }\textbf {\bibinfo {volume} {99}},\ \bibinfo
  {pages} {053606} (\bibinfo {year} {2019})}\BibitemShut {NoStop}%
\bibitem [{\citenamefont {Maffei}\ \emph {et~al.}(2018)\citenamefont {Maffei},
  \citenamefont {Dauphin}, \citenamefont {Cardano}, \citenamefont
  {Lewenstein},\ and\ \citenamefont {Massignan}}]{Maffei2018}%
  \BibitemOpen
  \bibfield  {author} {\bibinfo {author} {\bibfnamefont {M.}~\bibnamefont
  {Maffei}}, \bibinfo {author} {\bibfnamefont {A.}~\bibnamefont {Dauphin}},
  \bibinfo {author} {\bibfnamefont {F.}~\bibnamefont {Cardano}}, \bibinfo
  {author} {\bibfnamefont {M.}~\bibnamefont {Lewenstein}},\ and\ \bibinfo
  {author} {\bibfnamefont {P.}~\bibnamefont {Massignan}},\ }\href
  {https://doi.org/10.1088/1367-2630/aa9d4c} {\bibfield  {journal} {\bibinfo
  {journal} {New Journal of Physics}\ }\textbf {\bibinfo {volume} {20}},\
  \bibinfo {pages} {013023} (\bibinfo {year} {2018})}\BibitemShut {NoStop}%
\bibitem [{\citenamefont {Haller}\ \emph {et~al.}(2020)\citenamefont {Haller},
  \citenamefont {Massignan},\ and\ \citenamefont {Rizzi}}]{Haller2020}%
  \BibitemOpen
  \bibfield  {author} {\bibinfo {author} {\bibfnamefont {A.}~\bibnamefont
  {Haller}}, \bibinfo {author} {\bibfnamefont {P.}~\bibnamefont {Massignan}},\
  and\ \bibinfo {author} {\bibfnamefont {M.}~\bibnamefont {Rizzi}},\ }\href
  {https://doi.org/10.1103/PhysRevResearch.2.033200} {\bibfield  {journal}
  {\bibinfo  {journal} {Phys. Rev. Res.}\ }\textbf {\bibinfo {volume} {2}},\
  \bibinfo {pages} {033200} (\bibinfo {year} {2020})}\BibitemShut {NoStop}%
\bibitem [{\citenamefont {Kuno}\ \emph
  {et~al.}(2020{\natexlab{a}})\citenamefont {Kuno}, \citenamefont {Mizoguchi},\
  and\ \citenamefont {Hatsugai}}]{Kuno2020}%
  \BibitemOpen
  \bibfield  {author} {\bibinfo {author} {\bibfnamefont {Y.}~\bibnamefont
  {Kuno}}, \bibinfo {author} {\bibfnamefont {T.}~\bibnamefont {Mizoguchi}},\
  and\ \bibinfo {author} {\bibfnamefont {Y.}~\bibnamefont {Hatsugai}},\ }\href
  {https://doi.org/10.1103/PhysRevA.102.063325} {\bibfield  {journal} {\bibinfo
   {journal} {Phys. Rev. A}\ }\textbf {\bibinfo {volume} {102}},\ \bibinfo
  {pages} {063325} (\bibinfo {year} {2020}{\natexlab{a}})}\BibitemShut
  {NoStop}%
\bibitem [{\citenamefont {Mizoguchi}\ \emph
  {et~al.}(2021{\natexlab{a}})\citenamefont {Mizoguchi}, \citenamefont {Kuno},\
  and\ \citenamefont {Hatsugai}}]{Mizoguchi2021}%
  \BibitemOpen
  \bibfield  {author} {\bibinfo {author} {\bibfnamefont {T.}~\bibnamefont
  {Mizoguchi}}, \bibinfo {author} {\bibfnamefont {Y.}~\bibnamefont {Kuno}},\
  and\ \bibinfo {author} {\bibfnamefont {Y.}~\bibnamefont {Hatsugai}},\ }\href
  {https://doi.org/10.1103/PhysRevLett.126.016802} {\bibfield  {journal}
  {\bibinfo  {journal} {Phys. Rev. Lett.}\ }\textbf {\bibinfo {volume} {126}},\
  \bibinfo {pages} {016802} (\bibinfo {year} {2021}{\natexlab{a}})}\BibitemShut
  {NoStop}%
\bibitem [{\citenamefont {Heyl}\ \emph {et~al.}(2013)\citenamefont {Heyl},
  \citenamefont {Polkovnikov},\ and\ \citenamefont {Kehrein}}]{Heyl2013}%
  \BibitemOpen
  \bibfield  {author} {\bibinfo {author} {\bibfnamefont {M.}~\bibnamefont
  {Heyl}}, \bibinfo {author} {\bibfnamefont {A.}~\bibnamefont {Polkovnikov}},\
  and\ \bibinfo {author} {\bibfnamefont {S.}~\bibnamefont {Kehrein}},\ }\href
  {https://doi.org/10.1103/PhysRevLett.110.135704} {\bibfield  {journal}
  {\bibinfo  {journal} {Phys. Rev. Lett.}\ }\textbf {\bibinfo {volume} {110}},\
  \bibinfo {pages} {135704} (\bibinfo {year} {2013})}\BibitemShut {NoStop}%
\bibitem [{\citenamefont {Vajna}\ and\ \citenamefont
  {D\'ora}(2015)}]{Dora2015}%
  \BibitemOpen
  \bibfield  {author} {\bibinfo {author} {\bibfnamefont {S.}~\bibnamefont
  {Vajna}}\ and\ \bibinfo {author} {\bibfnamefont {B.}~\bibnamefont {D\'ora}},\
  }\href {https://doi.org/10.1103/PhysRevB.91.155127} {\bibfield  {journal}
  {\bibinfo  {journal} {Phys. Rev. B}\ }\textbf {\bibinfo {volume} {91}},\
  \bibinfo {pages} {155127} (\bibinfo {year} {2015})}\BibitemShut {NoStop}%
\bibitem [{\citenamefont {Okugawa}\ \emph {et~al.}(2021)\citenamefont
  {Okugawa}, \citenamefont {Oshiyama},\ and\ \citenamefont
  {Ohzeki}}]{Okugawa2021}%
  \BibitemOpen
  \bibfield  {author} {\bibinfo {author} {\bibfnamefont {R.}~\bibnamefont
  {Okugawa}}, \bibinfo {author} {\bibfnamefont {H.}~\bibnamefont {Oshiyama}},\
  and\ \bibinfo {author} {\bibfnamefont {M.}~\bibnamefont {Ohzeki}},\ }\href
  {https://doi.org/10.1103/PhysRevResearch.3.043064} {\bibfield  {journal}
  {\bibinfo  {journal} {Phys. Rev. Res.}\ }\textbf {\bibinfo {volume} {3}},\
  \bibinfo {pages} {043064} (\bibinfo {year} {2021})}\BibitemShut {NoStop}%
\bibitem [{\citenamefont {Jaksch}\ \emph {et~al.}(1998)\citenamefont {Jaksch},
  \citenamefont {Bruder}, \citenamefont {Cirac}, \citenamefont {Gardiner},\
  and\ \citenamefont {Zoller}}]{Jaksch1998}%
  \BibitemOpen
  \bibfield  {author} {\bibinfo {author} {\bibfnamefont {D.}~\bibnamefont
  {Jaksch}}, \bibinfo {author} {\bibfnamefont {C.}~\bibnamefont {Bruder}},
  \bibinfo {author} {\bibfnamefont {J.~I.}\ \bibnamefont {Cirac}}, \bibinfo
  {author} {\bibfnamefont {C.~W.}\ \bibnamefont {Gardiner}},\ and\ \bibinfo
  {author} {\bibfnamefont {P.}~\bibnamefont {Zoller}},\ }\href
  {https://doi.org/10.1103/PhysRevLett.81.3108} {\bibfield  {journal} {\bibinfo
   {journal} {Phys. Rev. Lett.}\ }\textbf {\bibinfo {volume} {81}},\ \bibinfo
  {pages} {3108} (\bibinfo {year} {1998})}\BibitemShut {NoStop}%
\bibitem [{\citenamefont {Jaksch}\ and\ \citenamefont
  {Zoller}(2003)}]{Jaksch2003}%
  \BibitemOpen
  \bibfield  {author} {\bibinfo {author} {\bibfnamefont {D.}~\bibnamefont
  {Jaksch}}\ and\ \bibinfo {author} {\bibfnamefont {P.}~\bibnamefont
  {Zoller}},\ }\href {https://doi.org/10.1088/1367-2630/5/1/356} {\bibfield
  {journal} {\bibinfo  {journal} {New Journal of Physics}\ }\textbf {\bibinfo
  {volume} {5}},\ \bibinfo {pages} {56} (\bibinfo {year} {2003})}\BibitemShut
  {NoStop}%
\bibitem [{\citenamefont {Lewenstein}\ \emph {et~al.}(2007)\citenamefont
  {Lewenstein}, \citenamefont {Sanpera}, \citenamefont {Ahufinger},
  \citenamefont {Damski}, \citenamefont {Sen(De)},\ and\ \citenamefont
  {Sen}}]{Maciej2007}%
  \BibitemOpen
  \bibfield  {author} {\bibinfo {author} {\bibfnamefont {M.}~\bibnamefont
  {Lewenstein}}, \bibinfo {author} {\bibfnamefont {A.}~\bibnamefont {Sanpera}},
  \bibinfo {author} {\bibfnamefont {V.}~\bibnamefont {Ahufinger}}, \bibinfo
  {author} {\bibfnamefont {B.}~\bibnamefont {Damski}}, \bibinfo {author}
  {\bibfnamefont {A.}~\bibnamefont {Sen(De)}},\ and\ \bibinfo {author}
  {\bibfnamefont {U.}~\bibnamefont {Sen}},\ }\href
  {https://doi.org/10.1080/00018730701223200} {\bibfield  {journal} {\bibinfo
  {journal} {Advances in Physics}\ }\textbf {\bibinfo {volume} {56}},\ \bibinfo
  {pages} {243} (\bibinfo {year} {2007})}\BibitemShut {NoStop}%
\bibitem [{\citenamefont {Bloch}\ \emph {et~al.}(2008)\citenamefont {Bloch},
  \citenamefont {Dalibard},\ and\ \citenamefont {Zwerger}}]{Bloch2008}%
  \BibitemOpen
  \bibfield  {author} {\bibinfo {author} {\bibfnamefont {I.}~\bibnamefont
  {Bloch}}, \bibinfo {author} {\bibfnamefont {J.}~\bibnamefont {Dalibard}},\
  and\ \bibinfo {author} {\bibfnamefont {W.}~\bibnamefont {Zwerger}},\ }\href
  {https://doi.org/10.1103/RevModPhys.80.885} {\bibfield  {journal} {\bibinfo
  {journal} {Rev. Mod. Phys.}\ }\textbf {\bibinfo {volume} {80}},\ \bibinfo
  {pages} {885} (\bibinfo {year} {2008})}\BibitemShut {NoStop}%
\bibitem [{\citenamefont {Joannopoulos}\ \emph {et~al.}(1997)\citenamefont
  {Joannopoulos}, \citenamefont {Villeneuve},\ and\ \citenamefont
  {Fan}}]{Joannopoulos1997}%
  \BibitemOpen
  \bibfield  {author} {\bibinfo {author} {\bibfnamefont {J.~D.}\ \bibnamefont
  {Joannopoulos}}, \bibinfo {author} {\bibfnamefont {P.~R.}\ \bibnamefont
  {Villeneuve}},\ and\ \bibinfo {author} {\bibfnamefont {S.}~\bibnamefont
  {Fan}},\ }\href {https://doi.org/10.1038/386143a0} {\bibfield  {journal}
  {\bibinfo  {journal} {Nature}\ }\textbf {\bibinfo {volume} {386}},\ \bibinfo
  {pages} {143} (\bibinfo {year} {1997})}\BibitemShut {NoStop}%
\bibitem [{\citenamefont {Ozawa}\ \emph {et~al.}(2019)\citenamefont {Ozawa},
  \citenamefont {Price}, \citenamefont {Amo}, \citenamefont {Goldman},
  \citenamefont {Hafezi}, \citenamefont {Lu}, \citenamefont {Rechtsman},
  \citenamefont {Schuster}, \citenamefont {Simon}, \citenamefont {Zilberberg},\
  and\ \citenamefont {Carusotto}}]{Ozawa2019}%
  \BibitemOpen
  \bibfield  {author} {\bibinfo {author} {\bibfnamefont {T.}~\bibnamefont
  {Ozawa}}, \bibinfo {author} {\bibfnamefont {H.~M.}\ \bibnamefont {Price}},
  \bibinfo {author} {\bibfnamefont {A.}~\bibnamefont {Amo}}, \bibinfo {author}
  {\bibfnamefont {N.}~\bibnamefont {Goldman}}, \bibinfo {author} {\bibfnamefont
  {M.}~\bibnamefont {Hafezi}}, \bibinfo {author} {\bibfnamefont
  {L.}~\bibnamefont {Lu}}, \bibinfo {author} {\bibfnamefont {M.~C.}\
  \bibnamefont {Rechtsman}}, \bibinfo {author} {\bibfnamefont {D.}~\bibnamefont
  {Schuster}}, \bibinfo {author} {\bibfnamefont {J.}~\bibnamefont {Simon}},
  \bibinfo {author} {\bibfnamefont {O.}~\bibnamefont {Zilberberg}},\ and\
  \bibinfo {author} {\bibfnamefont {I.}~\bibnamefont {Carusotto}},\ }\href
  {https://doi.org/10.1103/RevModPhys.91.015006} {\bibfield  {journal}
  {\bibinfo  {journal} {Rev. Mod. Phys.}\ }\textbf {\bibinfo {volume} {91}},\
  \bibinfo {pages} {015006} (\bibinfo {year} {2019})}\BibitemShut {NoStop}%
\bibitem [{\citenamefont {Kane}\ and\ \citenamefont
  {Lubensky}(2014)}]{Kane2014}%
  \BibitemOpen
  \bibfield  {author} {\bibinfo {author} {\bibfnamefont {C.~L.}\ \bibnamefont
  {Kane}}\ and\ \bibinfo {author} {\bibfnamefont {T.~C.}\ \bibnamefont
  {Lubensky}},\ }\href {https://doi.org/10.1038/nphys2835} {\bibfield
  {journal} {\bibinfo  {journal} {Nature Physics}\ }\textbf {\bibinfo {volume}
  {10}},\ \bibinfo {pages} {39} (\bibinfo {year} {2014})}\BibitemShut {NoStop}%
\bibitem [{\citenamefont {Ma}\ \emph {et~al.}(2019)\citenamefont {Ma},
  \citenamefont {Xiao},\ and\ \citenamefont {Chan}}]{Ma2019}%
  \BibitemOpen
  \bibfield  {author} {\bibinfo {author} {\bibfnamefont {G.}~\bibnamefont
  {Ma}}, \bibinfo {author} {\bibfnamefont {M.}~\bibnamefont {Xiao}},\ and\
  \bibinfo {author} {\bibfnamefont {C.~T.}\ \bibnamefont {Chan}},\ }\href
  {https://doi.org/10.1038/s42254-019-0030-x} {\bibfield  {journal} {\bibinfo
  {journal} {Nature Reviews Physics}\ }\textbf {\bibinfo {volume} {1}},\
  \bibinfo {pages} {281} (\bibinfo {year} {2019})}\BibitemShut {NoStop}%
\bibitem [{\citenamefont {Anderson}(1958)}]{Anderson1958}%
  \BibitemOpen
  \bibfield  {author} {\bibinfo {author} {\bibfnamefont {P.~W.}\ \bibnamefont
  {Anderson}},\ }\href {https://doi.org/10.1103/PhysRev.109.1492} {\bibfield
  {journal} {\bibinfo  {journal} {Phys. Rev.}\ }\textbf {\bibinfo {volume}
  {109}},\ \bibinfo {pages} {1492} (\bibinfo {year} {1958})}\BibitemShut
  {NoStop}%
\bibitem [{\citenamefont {Abrahams}\ \emph {et~al.}(1979)\citenamefont
  {Abrahams}, \citenamefont {Anderson}, \citenamefont {Licciardello},\ and\
  \citenamefont {Ramakrishnan}}]{Abrahams1979}%
  \BibitemOpen
  \bibfield  {author} {\bibinfo {author} {\bibfnamefont {E.}~\bibnamefont
  {Abrahams}}, \bibinfo {author} {\bibfnamefont {P.~W.}\ \bibnamefont
  {Anderson}}, \bibinfo {author} {\bibfnamefont {D.~C.}\ \bibnamefont
  {Licciardello}},\ and\ \bibinfo {author} {\bibfnamefont {T.~V.}\ \bibnamefont
  {Ramakrishnan}},\ }\href {https://doi.org/10.1103/PhysRevLett.42.673}
  {\bibfield  {journal} {\bibinfo  {journal} {Phys. Rev. Lett.}\ }\textbf
  {\bibinfo {volume} {42}},\ \bibinfo {pages} {673} (\bibinfo {year}
  {1979})}\BibitemShut {NoStop}%
\bibitem [{\citenamefont {Evers}\ and\ \citenamefont
  {Mirlin}(2008)}]{Evers2008}%
  \BibitemOpen
  \bibfield  {author} {\bibinfo {author} {\bibfnamefont {F.}~\bibnamefont
  {Evers}}\ and\ \bibinfo {author} {\bibfnamefont {A.~D.}\ \bibnamefont
  {Mirlin}},\ }\href {https://doi.org/10.1103/RevModPhys.80.1355} {\bibfield
  {journal} {\bibinfo  {journal} {Rev. Mod. Phys.}\ }\textbf {\bibinfo {volume}
  {80}},\ \bibinfo {pages} {1355} (\bibinfo {year} {2008})}\BibitemShut
  {NoStop}%
\bibitem [{\citenamefont {Wannier}(1960)}]{Wannier1960}%
  \BibitemOpen
  \bibfield  {author} {\bibinfo {author} {\bibfnamefont {G.~H.}\ \bibnamefont
  {Wannier}},\ }\href {https://doi.org/10.1103/PhysRev.117.432} {\bibfield
  {journal} {\bibinfo  {journal} {Phys. Rev.}\ }\textbf {\bibinfo {volume}
  {117}},\ \bibinfo {pages} {432} (\bibinfo {year} {1960})}\BibitemShut
  {NoStop}%
\bibitem [{\citenamefont {Atala}\ \emph {et~al.}(2013)\citenamefont {Atala},
  \citenamefont {Aidelsburger}, \citenamefont {Barreiro}, \citenamefont
  {Abanin}, \citenamefont {Kitagawa}, \citenamefont {Demler},\ and\
  \citenamefont {Bloch}}]{Atala2013}%
  \BibitemOpen
  \bibfield  {author} {\bibinfo {author} {\bibfnamefont {M.}~\bibnamefont
  {Atala}}, \bibinfo {author} {\bibfnamefont {M.}~\bibnamefont {Aidelsburger}},
  \bibinfo {author} {\bibfnamefont {J.~T.}\ \bibnamefont {Barreiro}}, \bibinfo
  {author} {\bibfnamefont {D.}~\bibnamefont {Abanin}}, \bibinfo {author}
  {\bibfnamefont {T.}~\bibnamefont {Kitagawa}}, \bibinfo {author}
  {\bibfnamefont {E.}~\bibnamefont {Demler}},\ and\ \bibinfo {author}
  {\bibfnamefont {I.}~\bibnamefont {Bloch}},\ }\href
  {https://doi.org/10.1038/nphys2790} {\bibfield  {journal} {\bibinfo
  {journal} {Nature Physics}\ }\textbf {\bibinfo {volume} {9}},\ \bibinfo
  {pages} {795} (\bibinfo {year} {2013})}\BibitemShut {NoStop}%
\bibitem [{\citenamefont {Kohlert}\ \emph {et~al.}(2023)\citenamefont
  {Kohlert}, \citenamefont {Scherg}, \citenamefont {Sala}, \citenamefont
  {Pollmann}, \citenamefont {Hebbe~Madhusudhana}, \citenamefont {Bloch},\ and\
  \citenamefont {Aidelsburger}}]{Kohlert2023}%
  \BibitemOpen
  \bibfield  {author} {\bibinfo {author} {\bibfnamefont {T.}~\bibnamefont
  {Kohlert}}, \bibinfo {author} {\bibfnamefont {S.}~\bibnamefont {Scherg}},
  \bibinfo {author} {\bibfnamefont {P.}~\bibnamefont {Sala}}, \bibinfo {author}
  {\bibfnamefont {F.}~\bibnamefont {Pollmann}}, \bibinfo {author}
  {\bibfnamefont {B.}~\bibnamefont {Hebbe~Madhusudhana}}, \bibinfo {author}
  {\bibfnamefont {I.}~\bibnamefont {Bloch}},\ and\ \bibinfo {author}
  {\bibfnamefont {M.}~\bibnamefont {Aidelsburger}},\ }\href
  {https://doi.org/10.1103/PhysRevLett.130.010201} {\bibfield  {journal}
  {\bibinfo  {journal} {Phys. Rev. Lett.}\ }\textbf {\bibinfo {volume} {130}},\
  \bibinfo {pages} {010201} (\bibinfo {year} {2023})}\BibitemShut {NoStop}%
\bibitem [{\citenamefont {Lim}\ \emph {et~al.}(2012)\citenamefont {Lim},
  \citenamefont {Fuchs},\ and\ \citenamefont {Montambaux}}]{Lim2012}%
  \BibitemOpen
  \bibfield  {author} {\bibinfo {author} {\bibfnamefont {L.-K.}\ \bibnamefont
  {Lim}}, \bibinfo {author} {\bibfnamefont {J.-N.}\ \bibnamefont {Fuchs}},\
  and\ \bibinfo {author} {\bibfnamefont {G.}~\bibnamefont {Montambaux}},\
  }\href {https://doi.org/10.1103/PhysRevLett.108.175303} {\bibfield  {journal}
  {\bibinfo  {journal} {Phys. Rev. Lett.}\ }\textbf {\bibinfo {volume} {108}},\
  \bibinfo {pages} {175303} (\bibinfo {year} {2012})}\BibitemShut {NoStop}%
\bibitem [{\citenamefont {Khomeriki}\ and\ \citenamefont
  {Flach}(2016)}]{Khomeriki2016}%
  \BibitemOpen
  \bibfield  {author} {\bibinfo {author} {\bibfnamefont {R.}~\bibnamefont
  {Khomeriki}}\ and\ \bibinfo {author} {\bibfnamefont {S.}~\bibnamefont
  {Flach}},\ }\href {https://doi.org/10.1103/PhysRevLett.116.245301} {\bibfield
   {journal} {\bibinfo  {journal} {Phys. Rev. Lett.}\ }\textbf {\bibinfo
  {volume} {116}},\ \bibinfo {pages} {245301} (\bibinfo {year}
  {2016})}\BibitemShut {NoStop}%
\bibitem [{\citenamefont {Di~Liberto}\ \emph {et~al.}(2020)\citenamefont
  {Di~Liberto}, \citenamefont {Goldman},\ and\ \citenamefont
  {Palumbo}}]{DiLiberto2020}%
  \BibitemOpen
  \bibfield  {author} {\bibinfo {author} {\bibfnamefont {M.}~\bibnamefont
  {Di~Liberto}}, \bibinfo {author} {\bibfnamefont {N.}~\bibnamefont
  {Goldman}},\ and\ \bibinfo {author} {\bibfnamefont {G.}~\bibnamefont
  {Palumbo}},\ }\href {https://doi.org/10.1038/s41467-020-19518-x} {\bibfield
  {journal} {\bibinfo  {journal} {Nature Communications}\ }\textbf {\bibinfo
  {volume} {11}},\ \bibinfo {pages} {5942} (\bibinfo {year}
  {2020})}\BibitemShut {NoStop}%
\bibitem [{\citenamefont {Kitamura}\ \emph {et~al.}(2020)\citenamefont
  {Kitamura}, \citenamefont {Nagaosa},\ and\ \citenamefont
  {Morimoto}}]{Kitamura2020}%
  \BibitemOpen
  \bibfield  {author} {\bibinfo {author} {\bibfnamefont {S.}~\bibnamefont
  {Kitamura}}, \bibinfo {author} {\bibfnamefont {N.}~\bibnamefont {Nagaosa}},\
  and\ \bibinfo {author} {\bibfnamefont {T.}~\bibnamefont {Morimoto}},\ }\href
  {https://doi.org/10.1038/s42005-020-0328-0} {\bibfield  {journal} {\bibinfo
  {journal} {Communications Physics}\ }\textbf {\bibinfo {volume} {3}},\
  \bibinfo {pages} {63} (\bibinfo {year} {2020})}\BibitemShut {NoStop}%
\bibitem [{\citenamefont {Vidal}\ \emph {et~al.}(1998)\citenamefont {Vidal},
  \citenamefont {Mosseri},\ and\ \citenamefont {Dou\ifmmode~\mbox{\c{c}}\else
  \c{c}\fi{}ot}}]{Vidal1998}%
  \BibitemOpen
  \bibfield  {author} {\bibinfo {author} {\bibfnamefont {J.}~\bibnamefont
  {Vidal}}, \bibinfo {author} {\bibfnamefont {R.}~\bibnamefont {Mosseri}},\
  and\ \bibinfo {author} {\bibfnamefont {B.}~\bibnamefont
  {Dou\ifmmode~\mbox{\c{c}}\else \c{c}\fi{}ot}},\ }\href
  {https://doi.org/10.1103/PhysRevLett.81.5888} {\bibfield  {journal} {\bibinfo
   {journal} {Phys. Rev. Lett.}\ }\textbf {\bibinfo {volume} {81}},\ \bibinfo
  {pages} {5888} (\bibinfo {year} {1998})}\BibitemShut {NoStop}%
\bibitem [{\citenamefont {Vidal}\ \emph {et~al.}(2000)\citenamefont {Vidal},
  \citenamefont {Dou\ifmmode~\mbox{\c{c}}\else \c{c}\fi{}ot}, \citenamefont
  {Mosseri},\ and\ \citenamefont {Butaud}}]{Vidal2000}%
  \BibitemOpen
  \bibfield  {author} {\bibinfo {author} {\bibfnamefont {J.}~\bibnamefont
  {Vidal}}, \bibinfo {author} {\bibfnamefont {B.}~\bibnamefont
  {Dou\ifmmode~\mbox{\c{c}}\else \c{c}\fi{}ot}}, \bibinfo {author}
  {\bibfnamefont {R.}~\bibnamefont {Mosseri}},\ and\ \bibinfo {author}
  {\bibfnamefont {P.}~\bibnamefont {Butaud}},\ }\href
  {https://doi.org/10.1103/PhysRevLett.85.3906} {\bibfield  {journal} {\bibinfo
   {journal} {Phys. Rev. Lett.}\ }\textbf {\bibinfo {volume} {85}},\ \bibinfo
  {pages} {3906} (\bibinfo {year} {2000})}\BibitemShut {NoStop}%
\bibitem [{\citenamefont {Vidal}\ \emph {et~al.}(2001)\citenamefont {Vidal},
  \citenamefont {Butaud}, \citenamefont {Dou\ifmmode~\mbox{\c{c}}\else
  \c{c}\fi{}ot},\ and\ \citenamefont {Mosseri}}]{Vidal2001}%
  \BibitemOpen
  \bibfield  {author} {\bibinfo {author} {\bibfnamefont {J.}~\bibnamefont
  {Vidal}}, \bibinfo {author} {\bibfnamefont {P.}~\bibnamefont {Butaud}},
  \bibinfo {author} {\bibfnamefont {B.}~\bibnamefont
  {Dou\ifmmode~\mbox{\c{c}}\else \c{c}\fi{}ot}},\ and\ \bibinfo {author}
  {\bibfnamefont {R.}~\bibnamefont {Mosseri}},\ }\href
  {https://doi.org/10.1103/PhysRevB.64.155306} {\bibfield  {journal} {\bibinfo
  {journal} {Phys. Rev. B}\ }\textbf {\bibinfo {volume} {64}},\ \bibinfo
  {pages} {155306} (\bibinfo {year} {2001})}\BibitemShut {NoStop}%
\bibitem [{\citenamefont {Dou\ifmmode~\mbox{\c{c}}\else \c{c}\fi{}ot}\ and\
  \citenamefont {Vidal}(2002)}]{Dousot2002}%
  \BibitemOpen
  \bibfield  {author} {\bibinfo {author} {\bibfnamefont {B.}~\bibnamefont
  {Dou\ifmmode~\mbox{\c{c}}\else \c{c}\fi{}ot}}\ and\ \bibinfo {author}
  {\bibfnamefont {J.}~\bibnamefont {Vidal}},\ }\href
  {https://doi.org/10.1103/PhysRevLett.88.227005} {\bibfield  {journal}
  {\bibinfo  {journal} {Phys. Rev. Lett.}\ }\textbf {\bibinfo {volume} {88}},\
  \bibinfo {pages} {227005} (\bibinfo {year} {2002})}\BibitemShut {NoStop}%
\bibitem [{\citenamefont {Mosseri}\ \emph {et~al.}(2022)\citenamefont
  {Mosseri}, \citenamefont {Vogeler},\ and\ \citenamefont
  {Vidal}}]{Mosseri2022}%
  \BibitemOpen
  \bibfield  {author} {\bibinfo {author} {\bibfnamefont {R.}~\bibnamefont
  {Mosseri}}, \bibinfo {author} {\bibfnamefont {R.}~\bibnamefont {Vogeler}},\
  and\ \bibinfo {author} {\bibfnamefont {J.}~\bibnamefont {Vidal}},\ }\href
  {https://doi.org/10.1103/PhysRevB.106.155120} {\bibfield  {journal} {\bibinfo
   {journal} {Phys. Rev. B}\ }\textbf {\bibinfo {volume} {106}},\ \bibinfo
  {pages} {155120} (\bibinfo {year} {2022})}\BibitemShut {NoStop}%
\bibitem [{\citenamefont {Ahmed}\ \emph {et~al.}(2022)\citenamefont {Ahmed},
  \citenamefont {Ramachandran}, \citenamefont {Khaymovich},\ and\ \citenamefont
  {Sharma}}]{Ahmed2022}%
  \BibitemOpen
  \bibfield  {author} {\bibinfo {author} {\bibfnamefont {A.}~\bibnamefont
  {Ahmed}}, \bibinfo {author} {\bibfnamefont {A.}~\bibnamefont {Ramachandran}},
  \bibinfo {author} {\bibfnamefont {I.~M.}\ \bibnamefont {Khaymovich}},\ and\
  \bibinfo {author} {\bibfnamefont {A.}~\bibnamefont {Sharma}},\ }\href
  {https://doi.org/10.1103/PhysRevB.106.205119} {\bibfield  {journal} {\bibinfo
   {journal} {Phys. Rev. B}\ }\textbf {\bibinfo {volume} {106}},\ \bibinfo
  {pages} {205119} (\bibinfo {year} {2022})}\BibitemShut {NoStop}%
\bibitem [{\citenamefont {Kolovsky}\ \emph {et~al.}(2023)\citenamefont
  {Kolovsky}, \citenamefont {Muraev},\ and\ \citenamefont
  {Flach}}]{Kolovsky2023}%
  \BibitemOpen
  \bibfield  {author} {\bibinfo {author} {\bibfnamefont {A.~R.}\ \bibnamefont
  {Kolovsky}}, \bibinfo {author} {\bibfnamefont {P.~S.}\ \bibnamefont
  {Muraev}},\ and\ \bibinfo {author} {\bibfnamefont {S.}~\bibnamefont {Flach}}}
  (\bibinfo {year} {2023}),\ \bibinfo {note} {arXiv:2303.00509}\BibitemShut
  {NoStop}%
\bibitem [{\citenamefont {Marques}\ \emph {et~al.}(2023)\citenamefont
  {Marques}, \citenamefont {M\"ogerle}, \citenamefont {Pelegr\'{\i}},
  \citenamefont {Flannigan}, \citenamefont {Dias},\ and\ \citenamefont
  {Daley}}]{Marques2023}%
  \BibitemOpen
  \bibfield  {author} {\bibinfo {author} {\bibfnamefont {A.~M.}\ \bibnamefont
  {Marques}}, \bibinfo {author} {\bibfnamefont {J.}~\bibnamefont {M\"ogerle}},
  \bibinfo {author} {\bibfnamefont {G.}~\bibnamefont {Pelegr\'{\i}}}, \bibinfo
  {author} {\bibfnamefont {S.}~\bibnamefont {Flannigan}}, \bibinfo {author}
  {\bibfnamefont {R.~G.}\ \bibnamefont {Dias}},\ and\ \bibinfo {author}
  {\bibfnamefont {A.~J.}\ \bibnamefont {Daley}},\ }\href
  {https://doi.org/10.1103/PhysRevResearch.5.023110} {\bibfield  {journal}
  {\bibinfo  {journal} {Phys. Rev. Res.}\ }\textbf {\bibinfo {volume} {5}},\
  \bibinfo {pages} {023110} (\bibinfo {year} {2023})}\BibitemShut {NoStop}%
\bibitem [{\citenamefont {Mukherjee}\ \emph {et~al.}(2018)\citenamefont
  {Mukherjee}, \citenamefont {Di~Liberto}, \citenamefont {\"Ohberg},
  \citenamefont {Thomson},\ and\ \citenamefont {Goldman}}]{Mukherjee2018}%
  \BibitemOpen
  \bibfield  {author} {\bibinfo {author} {\bibfnamefont {S.}~\bibnamefont
  {Mukherjee}}, \bibinfo {author} {\bibfnamefont {M.}~\bibnamefont
  {Di~Liberto}}, \bibinfo {author} {\bibfnamefont {P.}~\bibnamefont
  {\"Ohberg}}, \bibinfo {author} {\bibfnamefont {R.~R.}\ \bibnamefont
  {Thomson}},\ and\ \bibinfo {author} {\bibfnamefont {N.}~\bibnamefont
  {Goldman}},\ }\href {https://doi.org/10.1103/PhysRevLett.121.075502}
  {\bibfield  {journal} {\bibinfo  {journal} {Phys. Rev. Lett.}\ }\textbf
  {\bibinfo {volume} {121}},\ \bibinfo {pages} {075502} (\bibinfo {year}
  {2018})}\BibitemShut {NoStop}%
\bibitem [{\citenamefont {Kremer}\ \emph {et~al.}(2020)\citenamefont {Kremer},
  \citenamefont {Petrides}, \citenamefont {Meyer}, \citenamefont {Heinrich},
  \citenamefont {Zilberberg},\ and\ \citenamefont {Szameit}}]{Kremer2020}%
  \BibitemOpen
  \bibfield  {author} {\bibinfo {author} {\bibfnamefont {M.}~\bibnamefont
  {Kremer}}, \bibinfo {author} {\bibfnamefont {I.}~\bibnamefont {Petrides}},
  \bibinfo {author} {\bibfnamefont {E.}~\bibnamefont {Meyer}}, \bibinfo
  {author} {\bibfnamefont {M.}~\bibnamefont {Heinrich}}, \bibinfo {author}
  {\bibfnamefont {O.}~\bibnamefont {Zilberberg}},\ and\ \bibinfo {author}
  {\bibfnamefont {A.}~\bibnamefont {Szameit}},\ }\href
  {https://doi.org/10.1038/s41467-020-14692-4} {\bibfield  {journal} {\bibinfo
  {journal} {Nature Communications}\ }\textbf {\bibinfo {volume} {11}},\
  \bibinfo {pages} {907} (\bibinfo {year} {2020})}\BibitemShut {NoStop}%
\bibitem [{\citenamefont {Zhang}\ \emph {et~al.}(2023)\citenamefont {Zhang},
  \citenamefont {Wang}, \citenamefont {Sun},\ and\ \citenamefont
  {Zhang}}]{Zhang2023}%
  \BibitemOpen
  \bibfield  {author} {\bibinfo {author} {\bibfnamefont {W.}~\bibnamefont
  {Zhang}}, \bibinfo {author} {\bibfnamefont {H.}~\bibnamefont {Wang}},
  \bibinfo {author} {\bibfnamefont {H.}~\bibnamefont {Sun}},\ and\ \bibinfo
  {author} {\bibfnamefont {X.}~\bibnamefont {Zhang}},\ }\href
  {https://doi.org/10.1103/PhysRevLett.130.206401} {\bibfield  {journal}
  {\bibinfo  {journal} {Phys. Rev. Lett.}\ }\textbf {\bibinfo {volume} {130}},\
  \bibinfo {pages} {206401} (\bibinfo {year} {2023})}\BibitemShut {NoStop}%
\bibitem [{\citenamefont {Sutherland}(1986)}]{Sutherland1986}%
  \BibitemOpen
  \bibfield  {author} {\bibinfo {author} {\bibfnamefont {B.}~\bibnamefont
  {Sutherland}},\ }\href {https://doi.org/10.1103/PhysRevB.34.5208} {\bibfield
  {journal} {\bibinfo  {journal} {Phys. Rev. B}\ }\textbf {\bibinfo {volume}
  {34}},\ \bibinfo {pages} {5208} (\bibinfo {year} {1986})}\BibitemShut
  {NoStop}%
\bibitem [{\citenamefont {Lieb}(1989)}]{Lieb1989}%
  \BibitemOpen
  \bibfield  {author} {\bibinfo {author} {\bibfnamefont {E.~H.}\ \bibnamefont
  {Lieb}},\ }\href {https://doi.org/10.1103/PhysRevLett.62.1201} {\bibfield
  {journal} {\bibinfo  {journal} {Phys. Rev. Lett.}\ }\textbf {\bibinfo
  {volume} {62}},\ \bibinfo {pages} {1201} (\bibinfo {year}
  {1989})}\BibitemShut {NoStop}%
\bibitem [{\citenamefont {Brouwer}\ \emph {et~al.}(2002)\citenamefont
  {Brouwer}, \citenamefont {Racine}, \citenamefont {Furusaki}, \citenamefont
  {Hatsugai}, \citenamefont {Morita},\ and\ \citenamefont
  {Mudry}}]{Brouwer2002}%
  \BibitemOpen
  \bibfield  {author} {\bibinfo {author} {\bibfnamefont {P.~W.}\ \bibnamefont
  {Brouwer}}, \bibinfo {author} {\bibfnamefont {E.}~\bibnamefont {Racine}},
  \bibinfo {author} {\bibfnamefont {A.}~\bibnamefont {Furusaki}}, \bibinfo
  {author} {\bibfnamefont {Y.}~\bibnamefont {Hatsugai}}, \bibinfo {author}
  {\bibfnamefont {Y.}~\bibnamefont {Morita}},\ and\ \bibinfo {author}
  {\bibfnamefont {C.}~\bibnamefont {Mudry}},\ }\href
  {https://doi.org/10.1103/PhysRevB.66.014204} {\bibfield  {journal} {\bibinfo
  {journal} {Phys. Rev. B}\ }\textbf {\bibinfo {volume} {66}},\ \bibinfo
  {pages} {014204} (\bibinfo {year} {2002})}\BibitemShut {NoStop}%
\bibitem [{\citenamefont {Koshino}\ \emph {et~al.}(2014)\citenamefont
  {Koshino}, \citenamefont {Morimoto},\ and\ \citenamefont
  {Sato}}]{Koshino2014}%
  \BibitemOpen
  \bibfield  {author} {\bibinfo {author} {\bibfnamefont {M.}~\bibnamefont
  {Koshino}}, \bibinfo {author} {\bibfnamefont {T.}~\bibnamefont {Morimoto}},\
  and\ \bibinfo {author} {\bibfnamefont {M.}~\bibnamefont {Sato}},\ }\href
  {https://doi.org/10.1103/PhysRevB.90.115207} {\bibfield  {journal} {\bibinfo
  {journal} {Phys. Rev. B}\ }\textbf {\bibinfo {volume} {90}},\ \bibinfo
  {pages} {115207} (\bibinfo {year} {2014})}\BibitemShut {NoStop}%
\bibitem [{\citenamefont {McClure}(1956)}]{McClure1956}%
  \BibitemOpen
  \bibfield  {author} {\bibinfo {author} {\bibfnamefont {J.~W.}\ \bibnamefont
  {McClure}},\ }\href {https://doi.org/10.1103/PhysRev.104.666} {\bibfield
  {journal} {\bibinfo  {journal} {Phys. Rev.}\ }\textbf {\bibinfo {volume}
  {104}},\ \bibinfo {pages} {666} (\bibinfo {year} {1956})}\BibitemShut
  {NoStop}%
\bibitem [{\citenamefont {Arkinstall}\ \emph {et~al.}(2017)\citenamefont
  {Arkinstall}, \citenamefont {Teimourpour}, \citenamefont {Feng},
  \citenamefont {El-Ganainy},\ and\ \citenamefont
  {Schomerus}}]{Arkinstall2017}%
  \BibitemOpen
  \bibfield  {author} {\bibinfo {author} {\bibfnamefont {J.}~\bibnamefont
  {Arkinstall}}, \bibinfo {author} {\bibfnamefont {M.~H.}\ \bibnamefont
  {Teimourpour}}, \bibinfo {author} {\bibfnamefont {L.}~\bibnamefont {Feng}},
  \bibinfo {author} {\bibfnamefont {R.}~\bibnamefont {El-Ganainy}},\ and\
  \bibinfo {author} {\bibfnamefont {H.}~\bibnamefont {Schomerus}},\ }\href
  {https://doi.org/10.1103/PhysRevB.95.165109} {\bibfield  {journal} {\bibinfo
  {journal} {Phys. Rev. B}\ }\textbf {\bibinfo {volume} {95}},\ \bibinfo
  {pages} {165109} (\bibinfo {year} {2017})}\BibitemShut {NoStop}%
\bibitem [{\citenamefont {Attig}\ and\ \citenamefont
  {Trebst}(2017)}]{Attig2017}%
  \BibitemOpen
  \bibfield  {author} {\bibinfo {author} {\bibfnamefont {J.}~\bibnamefont
  {Attig}}\ and\ \bibinfo {author} {\bibfnamefont {S.}~\bibnamefont {Trebst}},\
  }\href {https://doi.org/10.1103/PhysRevB.96.085145} {\bibfield  {journal}
  {\bibinfo  {journal} {Phys. Rev. B}\ }\textbf {\bibinfo {volume} {96}},\
  \bibinfo {pages} {085145} (\bibinfo {year} {2017})}\BibitemShut {NoStop}%
\bibitem [{\citenamefont {Mizoguchi}\ \emph {et~al.}(2020)\citenamefont
  {Mizoguchi}, \citenamefont {Kuno},\ and\ \citenamefont
  {Hatsugai}}]{Mizoguchi2020_sq}%
  \BibitemOpen
  \bibfield  {author} {\bibinfo {author} {\bibfnamefont {T.}~\bibnamefont
  {Mizoguchi}}, \bibinfo {author} {\bibfnamefont {Y.}~\bibnamefont {Kuno}},\
  and\ \bibinfo {author} {\bibfnamefont {Y.}~\bibnamefont {Hatsugai}},\ }\href
  {https://doi.org/10.1103/PhysRevA.102.033527} {\bibfield  {journal} {\bibinfo
   {journal} {Phys. Rev. A}\ }\textbf {\bibinfo {volume} {102}},\ \bibinfo
  {pages} {033527} (\bibinfo {year} {2020})}\BibitemShut {NoStop}%
\bibitem [{\citenamefont {Mizoguchi}\ \emph
  {et~al.}(2021{\natexlab{b}})\citenamefont {Mizoguchi}, \citenamefont
  {Yoshida},\ and\ \citenamefont {Hatsugai}}]{Mizoguchi2021_SRTSM}%
  \BibitemOpen
  \bibfield  {author} {\bibinfo {author} {\bibfnamefont {T.}~\bibnamefont
  {Mizoguchi}}, \bibinfo {author} {\bibfnamefont {T.}~\bibnamefont {Yoshida}},\
  and\ \bibinfo {author} {\bibfnamefont {Y.}~\bibnamefont {Hatsugai}},\ }\href
  {https://doi.org/10.1103/PhysRevB.103.045136} {\bibfield  {journal} {\bibinfo
   {journal} {Phys. Rev. B}\ }\textbf {\bibinfo {volume} {103}},\ \bibinfo
  {pages} {045136} (\bibinfo {year} {2021}{\natexlab{b}})}\BibitemShut
  {NoStop}%
\bibitem [{\citenamefont {Yoshida}\ \emph {et~al.}(2021)\citenamefont
  {Yoshida}, \citenamefont {Mizoguchi}, \citenamefont {Kuno},\ and\
  \citenamefont {Hatsugai}}]{Yoshida2021}%
  \BibitemOpen
  \bibfield  {author} {\bibinfo {author} {\bibfnamefont {T.}~\bibnamefont
  {Yoshida}}, \bibinfo {author} {\bibfnamefont {T.}~\bibnamefont {Mizoguchi}},
  \bibinfo {author} {\bibfnamefont {Y.}~\bibnamefont {Kuno}},\ and\ \bibinfo
  {author} {\bibfnamefont {Y.}~\bibnamefont {Hatsugai}},\ }\href
  {https://doi.org/10.1103/PhysRevB.103.235130} {\bibfield  {journal} {\bibinfo
   {journal} {Phys. Rev. B}\ }\textbf {\bibinfo {volume} {103}},\ \bibinfo
  {pages} {235130} (\bibinfo {year} {2021})}\BibitemShut {NoStop}%
\bibitem [{\citenamefont {Navarro-Labastida}\ and\ \citenamefont
  {Naumis}(2023)}]{Navarro2023}%
  \BibitemOpen
  \bibfield  {author} {\bibinfo {author} {\bibfnamefont {L.~A.}\ \bibnamefont
  {Navarro-Labastida}}\ and\ \bibinfo {author} {\bibfnamefont {G.~G.}\
  \bibnamefont {Naumis}},\ }\href {https://doi.org/10.1103/PhysRevB.107.155428}
  {\bibfield  {journal} {\bibinfo  {journal} {Phys. Rev. B}\ }\textbf {\bibinfo
  {volume} {107}},\ \bibinfo {pages} {155428} (\bibinfo {year}
  {2023})}\BibitemShut {NoStop}%
\bibitem [{\citenamefont {Matsumoto}\ \emph {et~al.}(2023)\citenamefont
  {Matsumoto}, \citenamefont {Mizoguchi},\ and\ \citenamefont
  {Hatsugai}}]{Matsumoto2023}%
  \BibitemOpen
  \bibfield  {author} {\bibinfo {author} {\bibfnamefont {D.}~\bibnamefont
  {Matsumoto}}, \bibinfo {author} {\bibfnamefont {T.}~\bibnamefont
  {Mizoguchi}},\ and\ \bibinfo {author} {\bibfnamefont {Y.}~\bibnamefont
  {Hatsugai}},\ }\href {https://doi.org/10.7566/JPSJ.92.034705} {\bibfield
  {journal} {\bibinfo  {journal} {Journal of the Physical Society of Japan}\
  }\textbf {\bibinfo {volume} {92}},\ \bibinfo {pages} {034705} (\bibinfo
  {year} {2023})}\BibitemShut {NoStop}%
\bibitem [{\citenamefont {Mizoguchi}\ and\ \citenamefont
  {Hatsugai}(2023)}]{Mizoguchi2023}%
  \BibitemOpen
  \bibfield  {author} {\bibinfo {author} {\bibfnamefont {T.}~\bibnamefont
  {Mizoguchi}}\ and\ \bibinfo {author} {\bibfnamefont {Y.}~\bibnamefont
  {Hatsugai}},\ }\href {https://doi.org/10.1103/PhysRevB.107.094201} {\bibfield
   {journal} {\bibinfo  {journal} {Phys. Rev. B}\ }\textbf {\bibinfo {volume}
  {107}},\ \bibinfo {pages} {094201} (\bibinfo {year} {2023})}\BibitemShut
  {NoStop}%
\bibitem [{rem()}]{remark}%
  \BibitemOpen
  \href@noop {} {}\bibinfo {note} {Note that, when $L$ is even, there is no
  plaquette exactly satisfying $\Phi_n = \pi$. Specifically, $\Phi_{L/2} = \pi
  \cdot \frac{L-2}{L-1}$ and $\Phi_{L/2} = \pi \cdot
  \frac{L}{L-1}$.}\BibitemShut {Stop}%
\bibitem [{\citenamefont {Creutz}(1999)}]{Creutz1999}%
  \BibitemOpen
  \bibfield  {author} {\bibinfo {author} {\bibfnamefont {M.}~\bibnamefont
  {Creutz}},\ }\href {https://doi.org/10.1103/PhysRevLett.83.2636} {\bibfield
  {journal} {\bibinfo  {journal} {Phys. Rev. Lett.}\ }\textbf {\bibinfo
  {volume} {83}},\ \bibinfo {pages} {2636} (\bibinfo {year}
  {1999})}\BibitemShut {NoStop}%
\bibitem [{\citenamefont {Creutz}(2001)}]{Creutz2001}%
  \BibitemOpen
  \bibfield  {author} {\bibinfo {author} {\bibfnamefont {M.}~\bibnamefont
  {Creutz}},\ }\href {https://doi.org/10.1103/RevModPhys.73.119} {\bibfield
  {journal} {\bibinfo  {journal} {Rev. Mod. Phys.}\ }\textbf {\bibinfo {volume}
  {73}},\ \bibinfo {pages} {119} (\bibinfo {year} {2001})}\BibitemShut
  {NoStop}%
\bibitem [{\citenamefont {Kuno}\ \emph
  {et~al.}(2020{\natexlab{b}})\citenamefont {Kuno}, \citenamefont {Orito},\
  and\ \citenamefont {Ichinose}}]{Kuno2020_C}%
  \BibitemOpen
  \bibfield  {author} {\bibinfo {author} {\bibfnamefont {Y.}~\bibnamefont
  {Kuno}}, \bibinfo {author} {\bibfnamefont {T.}~\bibnamefont {Orito}},\ and\
  \bibinfo {author} {\bibfnamefont {I.}~\bibnamefont {Ichinose}},\ }\href
  {https://doi.org/10.1088/1367-2630/ab6352} {\bibfield  {journal} {\bibinfo
  {journal} {New Journal of Physics}\ }\textbf {\bibinfo {volume} {22}},\
  \bibinfo {pages} {013032} (\bibinfo {year} {2020}{\natexlab{b}})}\BibitemShut
  {NoStop}%
\bibitem [{\citenamefont {Cheneau}\ \emph {et~al.}(2012)\citenamefont
  {Cheneau}, \citenamefont {Barmettler}, \citenamefont {Poletti}, \citenamefont
  {Endres}, \citenamefont {Schau{\ss}}, \citenamefont {Fukuhara}, \citenamefont
  {Gross}, \citenamefont {Bloch}, \citenamefont {Kollath},\ and\ \citenamefont
  {Kuhr}}]{Cheneau2012}%
  \BibitemOpen
  \bibfield  {author} {\bibinfo {author} {\bibfnamefont {M.}~\bibnamefont
  {Cheneau}}, \bibinfo {author} {\bibfnamefont {P.}~\bibnamefont {Barmettler}},
  \bibinfo {author} {\bibfnamefont {D.}~\bibnamefont {Poletti}}, \bibinfo
  {author} {\bibfnamefont {M.}~\bibnamefont {Endres}}, \bibinfo {author}
  {\bibfnamefont {P.}~\bibnamefont {Schau{\ss}}}, \bibinfo {author}
  {\bibfnamefont {T.}~\bibnamefont {Fukuhara}}, \bibinfo {author}
  {\bibfnamefont {C.}~\bibnamefont {Gross}}, \bibinfo {author} {\bibfnamefont
  {I.}~\bibnamefont {Bloch}}, \bibinfo {author} {\bibfnamefont
  {C.}~\bibnamefont {Kollath}},\ and\ \bibinfo {author} {\bibfnamefont
  {S.}~\bibnamefont {Kuhr}},\ }\href {https://doi.org/10.1038/nature10748}
  {\bibfield  {journal} {\bibinfo  {journal} {Nature}\ }\textbf {\bibinfo
  {volume} {481}},\ \bibinfo {pages} {484} (\bibinfo {year}
  {2012})}\BibitemShut {NoStop}%
\bibitem [{\citenamefont {Ronzheimer}\ \emph {et~al.}(2013)\citenamefont
  {Ronzheimer}, \citenamefont {Schreiber}, \citenamefont {Braun}, \citenamefont
  {Hodgman}, \citenamefont {Langer}, \citenamefont {McCulloch}, \citenamefont
  {Heidrich-Meisner}, \citenamefont {Bloch},\ and\ \citenamefont
  {Schneider}}]{Ronzheimer2013}%
  \BibitemOpen
  \bibfield  {author} {\bibinfo {author} {\bibfnamefont {J.~P.}\ \bibnamefont
  {Ronzheimer}}, \bibinfo {author} {\bibfnamefont {M.}~\bibnamefont
  {Schreiber}}, \bibinfo {author} {\bibfnamefont {S.}~\bibnamefont {Braun}},
  \bibinfo {author} {\bibfnamefont {S.~S.}\ \bibnamefont {Hodgman}}, \bibinfo
  {author} {\bibfnamefont {S.}~\bibnamefont {Langer}}, \bibinfo {author}
  {\bibfnamefont {I.~P.}\ \bibnamefont {McCulloch}}, \bibinfo {author}
  {\bibfnamefont {F.}~\bibnamefont {Heidrich-Meisner}}, \bibinfo {author}
  {\bibfnamefont {I.}~\bibnamefont {Bloch}},\ and\ \bibinfo {author}
  {\bibfnamefont {U.}~\bibnamefont {Schneider}},\ }\href
  {https://doi.org/10.1103/PhysRevLett.110.205301} {\bibfield  {journal}
  {\bibinfo  {journal} {Phys. Rev. Lett.}\ }\textbf {\bibinfo {volume} {110}},\
  \bibinfo {pages} {205301} (\bibinfo {year} {2013})}\BibitemShut {NoStop}%
\bibitem [{\citenamefont {Jreissaty}\ \emph {et~al.}(2013)\citenamefont
  {Jreissaty}, \citenamefont {Carrasquilla},\ and\ \citenamefont
  {Rigol}}]{Jreissaty2013}%
  \BibitemOpen
  \bibfield  {author} {\bibinfo {author} {\bibfnamefont {A.}~\bibnamefont
  {Jreissaty}}, \bibinfo {author} {\bibfnamefont {J.}~\bibnamefont
  {Carrasquilla}},\ and\ \bibinfo {author} {\bibfnamefont {M.}~\bibnamefont
  {Rigol}},\ }\href {https://doi.org/10.1103/PhysRevA.88.031606} {\bibfield
  {journal} {\bibinfo  {journal} {Phys. Rev. A}\ }\textbf {\bibinfo {volume}
  {88}},\ \bibinfo {pages} {031606} (\bibinfo {year} {2013})}\BibitemShut
  {NoStop}%
\bibitem [{\citenamefont {Vidmar}\ \emph {et~al.}(2013)\citenamefont {Vidmar},
  \citenamefont {Langer}, \citenamefont {McCulloch}, \citenamefont {Schneider},
  \citenamefont {Schollw\"ock},\ and\ \citenamefont
  {Heidrich-Meisner}}]{Vidmar2013}%
  \BibitemOpen
  \bibfield  {author} {\bibinfo {author} {\bibfnamefont {L.}~\bibnamefont
  {Vidmar}}, \bibinfo {author} {\bibfnamefont {S.}~\bibnamefont {Langer}},
  \bibinfo {author} {\bibfnamefont {I.~P.}\ \bibnamefont {McCulloch}}, \bibinfo
  {author} {\bibfnamefont {U.}~\bibnamefont {Schneider}}, \bibinfo {author}
  {\bibfnamefont {U.}~\bibnamefont {Schollw\"ock}},\ and\ \bibinfo {author}
  {\bibfnamefont {F.}~\bibnamefont {Heidrich-Meisner}},\ }\href
  {https://doi.org/10.1103/PhysRevB.88.235117} {\bibfield  {journal} {\bibinfo
  {journal} {Phys. Rev. B}\ }\textbf {\bibinfo {volume} {88}},\ \bibinfo
  {pages} {235117} (\bibinfo {year} {2013})}\BibitemShut {NoStop}%
\bibitem [{\citenamefont {Aidelsburger}\ \emph {et~al.}(2011)\citenamefont
  {Aidelsburger}, \citenamefont {Atala}, \citenamefont {Nascimb\`ene},
  \citenamefont {Trotzky}, \citenamefont {Chen},\ and\ \citenamefont
  {Bloch}}]{Aidelsburger2011}%
  \BibitemOpen
  \bibfield  {author} {\bibinfo {author} {\bibfnamefont {M.}~\bibnamefont
  {Aidelsburger}}, \bibinfo {author} {\bibfnamefont {M.}~\bibnamefont {Atala}},
  \bibinfo {author} {\bibfnamefont {S.}~\bibnamefont {Nascimb\`ene}}, \bibinfo
  {author} {\bibfnamefont {S.}~\bibnamefont {Trotzky}}, \bibinfo {author}
  {\bibfnamefont {Y.-A.}\ \bibnamefont {Chen}},\ and\ \bibinfo {author}
  {\bibfnamefont {I.}~\bibnamefont {Bloch}},\ }\href
  {https://doi.org/10.1103/PhysRevLett.107.255301} {\bibfield  {journal}
  {\bibinfo  {journal} {Phys. Rev. Lett.}\ }\textbf {\bibinfo {volume} {107}},\
  \bibinfo {pages} {255301} (\bibinfo {year} {2011})}\BibitemShut {NoStop}%
\end{thebibliography}%

\end{document}